\documentclass[aps,prb,twocolumn,floatfix,showpacs,groupedaddress,superscriptaddress]{revtex4-1}

\usepackage{graphicx}  
\usepackage{dcolumn}   
\usepackage{bm}        
\usepackage{amssymb}   
\usepackage{amsmath}
\usepackage{braket}
\usepackage[mathscr]{euscript}
\usepackage{xcolor}
\usepackage[caption=false,labelformat=simple]{subfig}

\begin{document}

\title{Quantum Hall studies of a Semi-Dirac Nanoribbon}
\author{Priyanka Sinha}
\email{sinhapriyanka2016@iitg.ac.in}
\affiliation{Department of Physics, Indian Institute of Technology Guwahati\\ Guwahati-781039, Assam, India}
\author{Shuichi Murakami}
\email{murakami@stat.phys.titech.ac.jp}
\affiliation{Department of Physics, Tokyo Institute of Technology\\
2-12-1 Ookayama, Meguro-ku, Tokyo 152-8551, Japan}
\author{Saurabh Basu}
\email{saurabh@iitg.ac.in}
\affiliation{Department of Physics, Indian Institute of Technology Guwahati\\ Guwahati-781039, Assam, India}
\date{\today}
\begin{abstract}
Here we comprehensively investigate Landau levels, Hofstadter butterfly and transport properties of a semi-Dirac nanoribbon
in a perpendicular magnetic field using a recently developed real-space implementation of the Kubo formula based on Kernel Polynomial Method. 
A Dirac ribbon is considered to compare and contrast our results for a semi-Dirac system. We find that the Landau levels being non-equidistant from
each other for the semi-Dirac case (true for a Dirac as well), the flatness of the energy bands vanishes in the bulk and becomes dispersive
for a semi-Dirac ribbon in contrast to a Dirac system. This feature is most discernible for intermediate values of the external field. We further compute the 
longitudinal ($\sigma_{xx}$ and $\sigma_{yy}$) and the transverse or Hall ($\sigma_{xy}$) conductivities where the Hall conductivity 
shows a familiar quantization, namely, $\sigma_{xy} \propto 2n$ (the factor `2' includes the spin degeneracy) which is highly distinct
from a Dirac system, such as graphene. We also observe anisotropic behavior in magneto-transport in a semi-Dirac ribbon
owing to the dispersion anomalies in two different longitudinal directions. Our studies may have important ramifications for monolayer phosphorene. 

\end{abstract}
%%
%\pacs{72.80.Vp, 73.20.At, 73.22.Gk,}
\maketitle
%%
%%
%%-----------------------------------INTRODUCTION----------------------------------------------

\section{Introduction}
In the past few decades, graphene has attracted much attention due to its peculiar dispersion relation at low energies, similar to the spectrum of relativistic particles described by the Dirac theory\cite{wallace,neto}. More precisely, the spectrum has two cones, the so-called “Dirac cones” in the vicinity of two non-equivalent points $K_1$ and $K_2$ in the reciprocal space. Anisotropy in graphene was another interesting aspect which was discussed long ago by Pauling\cite{pauling}, and could be induced by uniaxial stress or bending of a graphene sheet. The main motive is to tune the hopping energy between neighbouring carbon atoms with precision, which was later found to be feasible in optical lattices \cite{tarruell} via controlling the lattice potential in order to have a handle on the effective mass in a honeycomb lattice. In a tight-binding model for graphene, if one of the three nearest-neighbor hopping energies is tuned, the two Dirac points with opposite chiralities approach each other and merge into one forming the so-called semi-Dirac point. The band dispersion simultaneously exhibits massless Dirac (linear) and massive fermionic (quadratic) features along two different directions, thereby producing a highly anisotropic electronic dispersion\cite{dietl,pickett}. The materials that host such anisotropic dispersion are phosphorene under pressure and doping \cite{rodin,guan}, electric fields \cite{rudenko,dut}, TiO2/VO2 superlattices \cite{pickett, pardo1, pardo2}, graphene under deformation \cite{mon}, and BEDT-TTF$_{2}$I$_{3}$ salt under pressure \cite{konno,pichon}. Experimentally, semi-Dirac dispersion has been observed in a few-layer black phosphorene by means of the in situ deposition of potassium atoms \cite{kim}. A straightforward approach to realize semi-Dirac materials can be achieved by breaking the hexagonal symmetry of the honeycomb lattice, e.g, by strain. However, directly applying strain to realize the transition in materials, such as graphene or silicene is prohibited by the exorbitant magnitude of the strain required, which would eventually disintegrate them \cite{si,zhao}. Some successfully synthesized graphene-like honeycomb materials, such as silicene \cite{vogt,meng}, germanene \cite{li} and stanene \cite{zhu}, are found to be easily oxidized or they chemically absorb other atoms because of their buckling geometries \cite{molle, spencer, houselt, ciraci}. It is these absorbed atoms that will modify the hopping energies in the honeycomb lattice which is essentially applying a strain that creates a differential hopping.\par 

\begin{figure}[h]
\begin{center}
\subfloat{\includegraphics[width=0.45\textwidth]{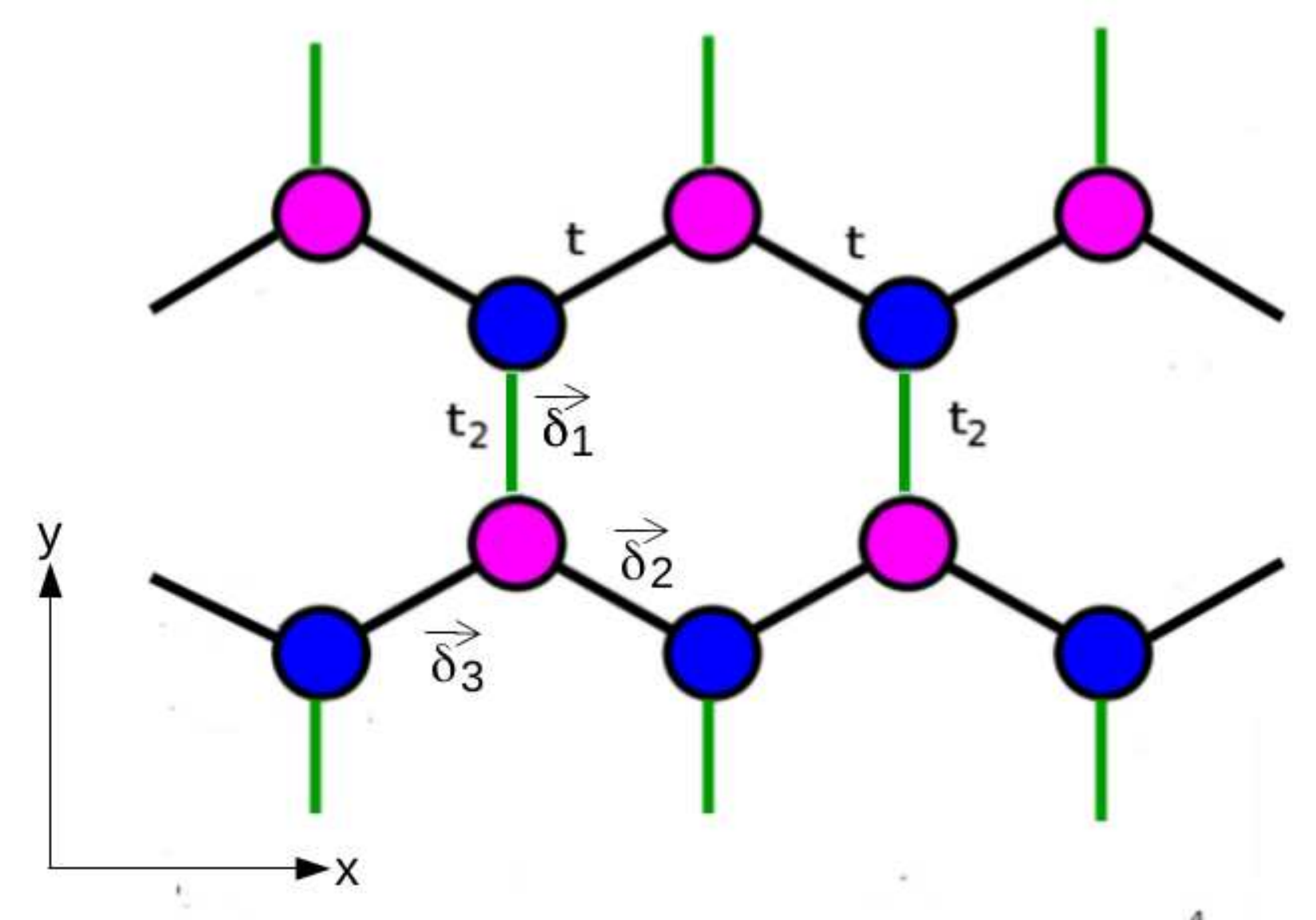}\label{fig:1}}
\caption{(Color online) Schematic diagram of hexagonal lattice geometry of a semi-Dirac system with different hopping parameters $t$ and $t_2$ is shown. The black corresponds to hopping, $t$ whereas the green corresponds to $t_2$. Two sublattices are denoted by two different colors (blue and magenta). $\vec{\delta_1}$, $\vec{\delta_2}$ and $\vec{\delta_3}$ are the nearest neighbor real space vectors.}
\label{fig:1}
\end{center}
\end{figure}
On the other hand, behaviors of electrons in graphene, exposed to a strong perpendicular magnetic field played an important role not only for the discovery of quantum Hall effects\cite{vp,zheng}, but also for proving the existence of massless Dirac particles\cite{firsov,zhang}. The unconventional Hall conductivity was found to be quantized as $\sigma_{xy}=2(2n+1) e^2/h$\cite{firsov,zhang}, where both the spin and the valley degeneracies are taken into account. Experimental measurements confirm that the Landau levels of a monolayer graphene obey the relation, $E_n=sgn(n)\sqrt{2 \hbar v_{F}e|n|B}$, where $v_{F}$=$10^6$ m/s is the Fermi velocity, $B$ is the magnetic field and $n$ denote Landau level indices\cite{sadowski1,sadowski2}. 

Recently, quite a few studies on Landau levels and transport properties in presence of a magnetic field in phosphorene have been reported \cite{chang,roldan}. More precisely, they have found that the anisotropic band structure that leads to Hall quantization in presence of a perpendicular magnetic field is similar to that of a conventional two-dimensional electron gas (2DEG). Since phosphorene may be considered as a realistic material that possesses semi-Dirac properties, it is necessary to pursue quantum Hall studies on the semi-Dirac systems. As discussed above, the energy dispersion of phosphorene is similar to that of the semi-Dirac systems, it is likely that other properties too show similar characteristics.\par   
In this work, we have explored the influence of magnetic field for a semi-Dirac system using a tight-binding Hamiltonian on a honeycomb lattice. We study the Landau level spectrum and Hofstadter butterfly using a nanoribbon in order to show that the semi-Dirac system has quite distinct properties as compared to Dirac fermions. We also calculate the density of states (DOS) via the tight-binding propagation method\cite{yuan,yuan2}, which is a sophisticated numerical tool used in large-scale calculations for any realistic system. We have implemented the recently developed real-space order-$N$ quantum transport approach to calculate the Kubo conductivities as a function of the Fermi energy for moderate as well as very high values of the magnetic field\cite{rappoprt}. The Hall conductivity in a semi-Dirac system shows the {\it {standard}} quantization, namely, $\sigma_{xy}\propto 2n$ as compared to the previously observed {\it {anomalous}} quantization, that is, $\sigma_{xy}\propto 4(n+1/2)$ for a Dirac system. The longitudinal conductivities show highly anisotropic behavior in one direction compared to the other, which is obviously absent for Dirac systems.\par 
The paper is organized as follows. The low energy tight-binding Hamiltonian is described in sec.~\ref{II}. We have further studied the Landau level spectra and the Hofstadter butterfly for a nanoribbon in presence of a magnetic field in sec.~\ref{III}. The transport properties are investigated by computing the Hall and the longitudinal conductivities in sec.~\ref{IV}. We conclude with a brief summary in sec.~\ref{V}. 

\section{Model Hamiltonian}\label{II}
\begin{figure*}[!ht!]
\begin{center}
\subfloat[]{\includegraphics[width=0.37\textwidth]{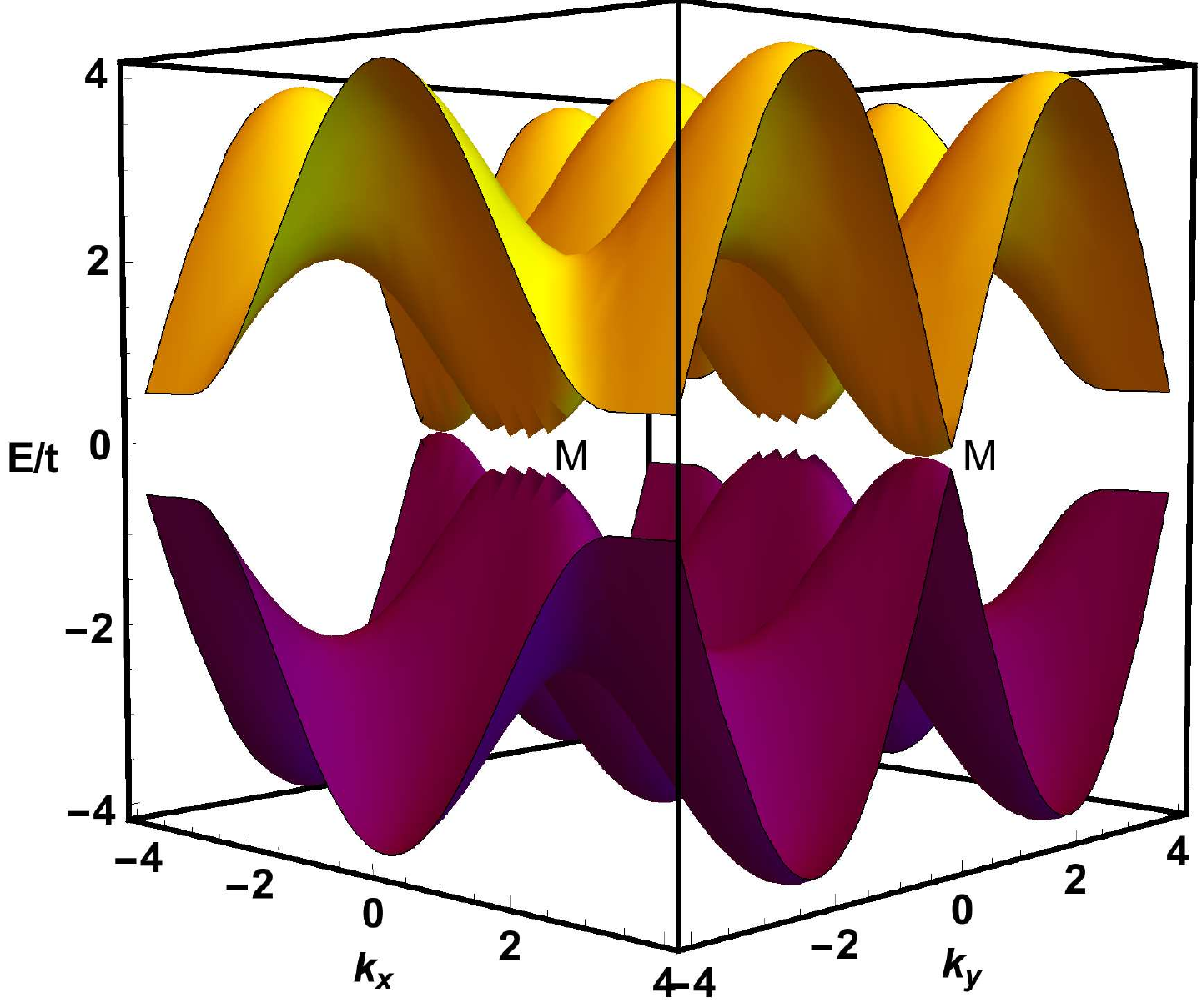}\label{fig:2a}} \hspace*{2.5 cm}
\subfloat[]{\includegraphics[width=0.38\textwidth]{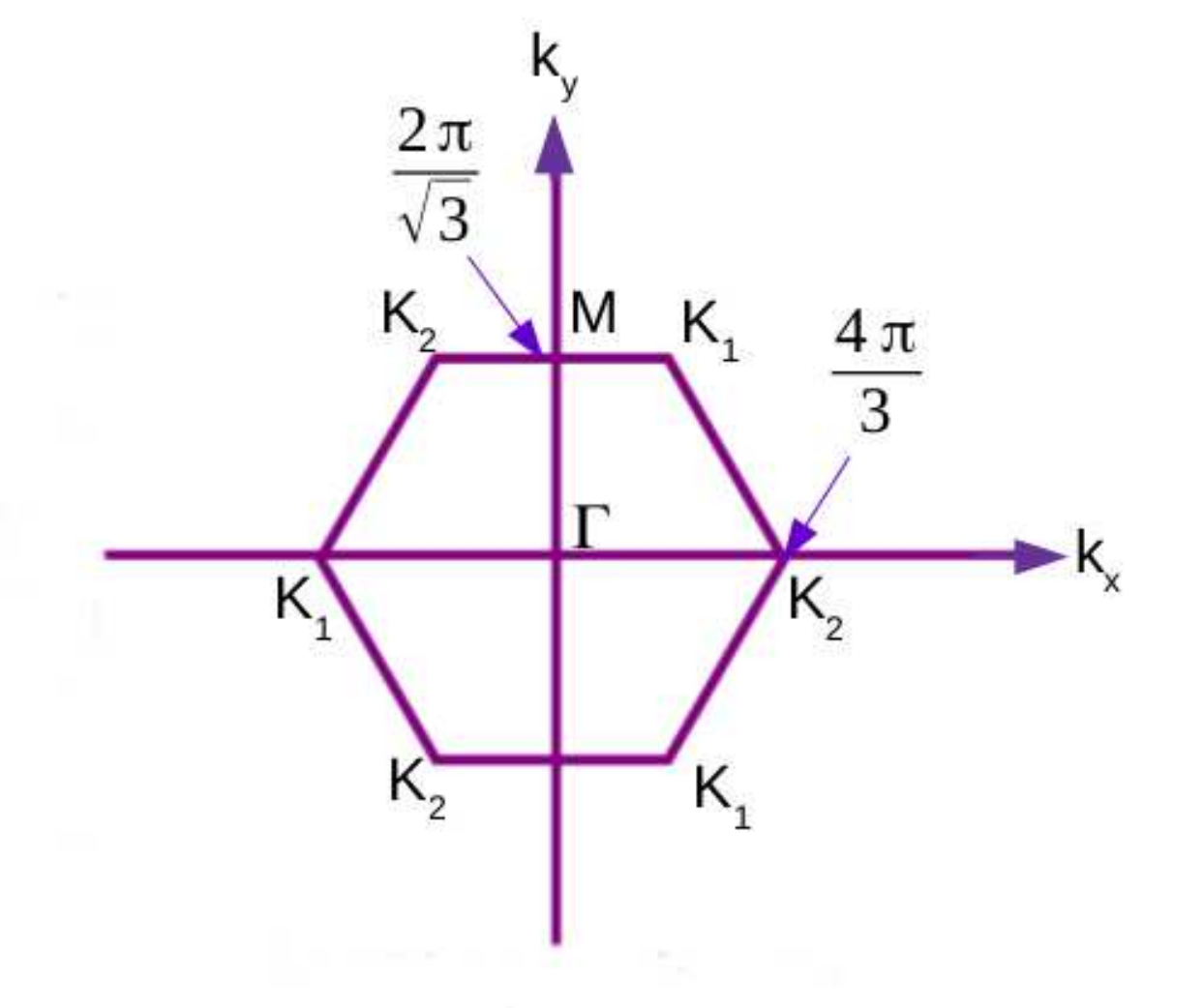}\label{fig:2b}}\\
\subfloat[]{\includegraphics[width=0.35\textwidth]{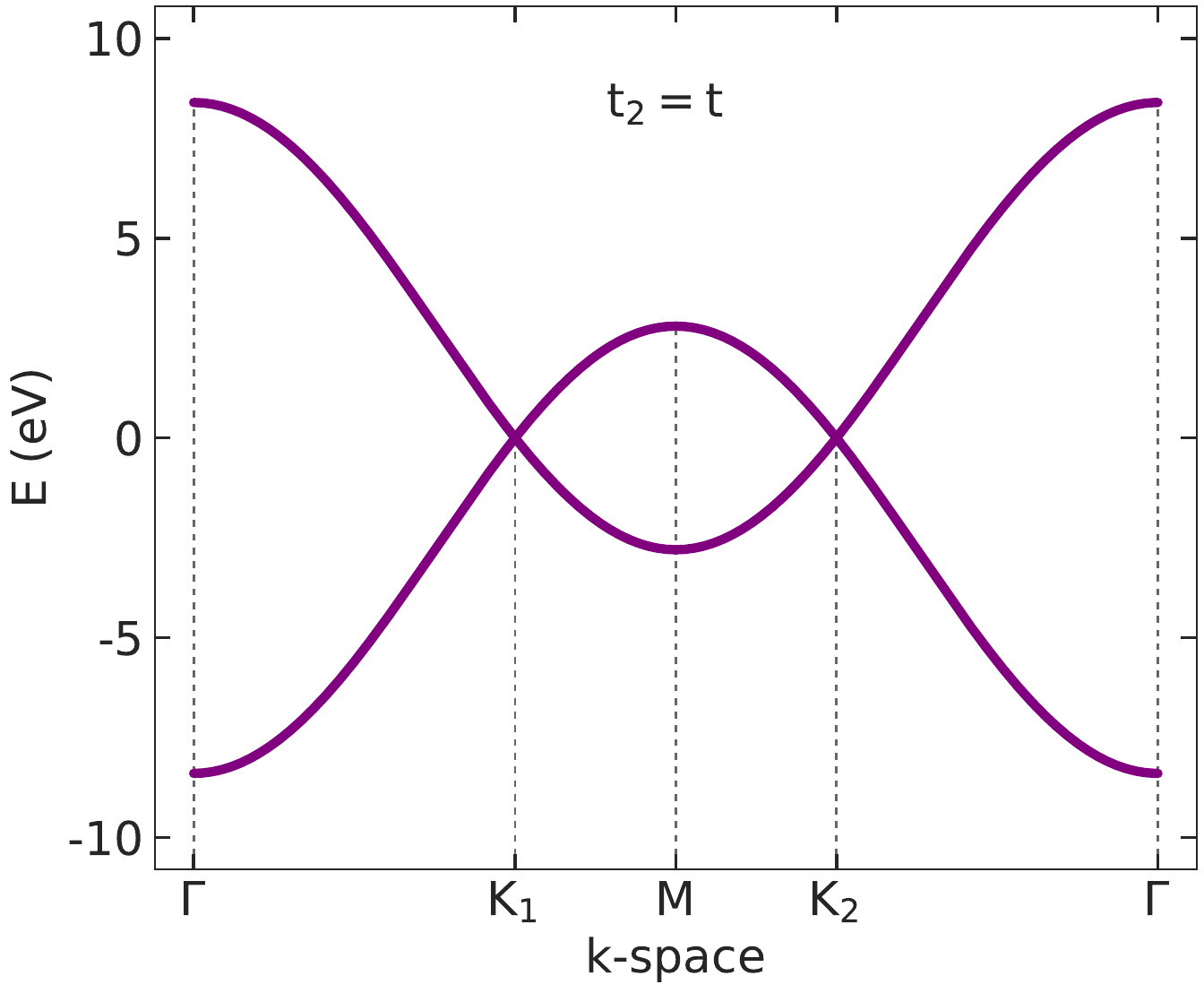}\label{fig:2c}} \hspace*{2.5 cm}
\subfloat[]{\includegraphics[width=0.35\textwidth]{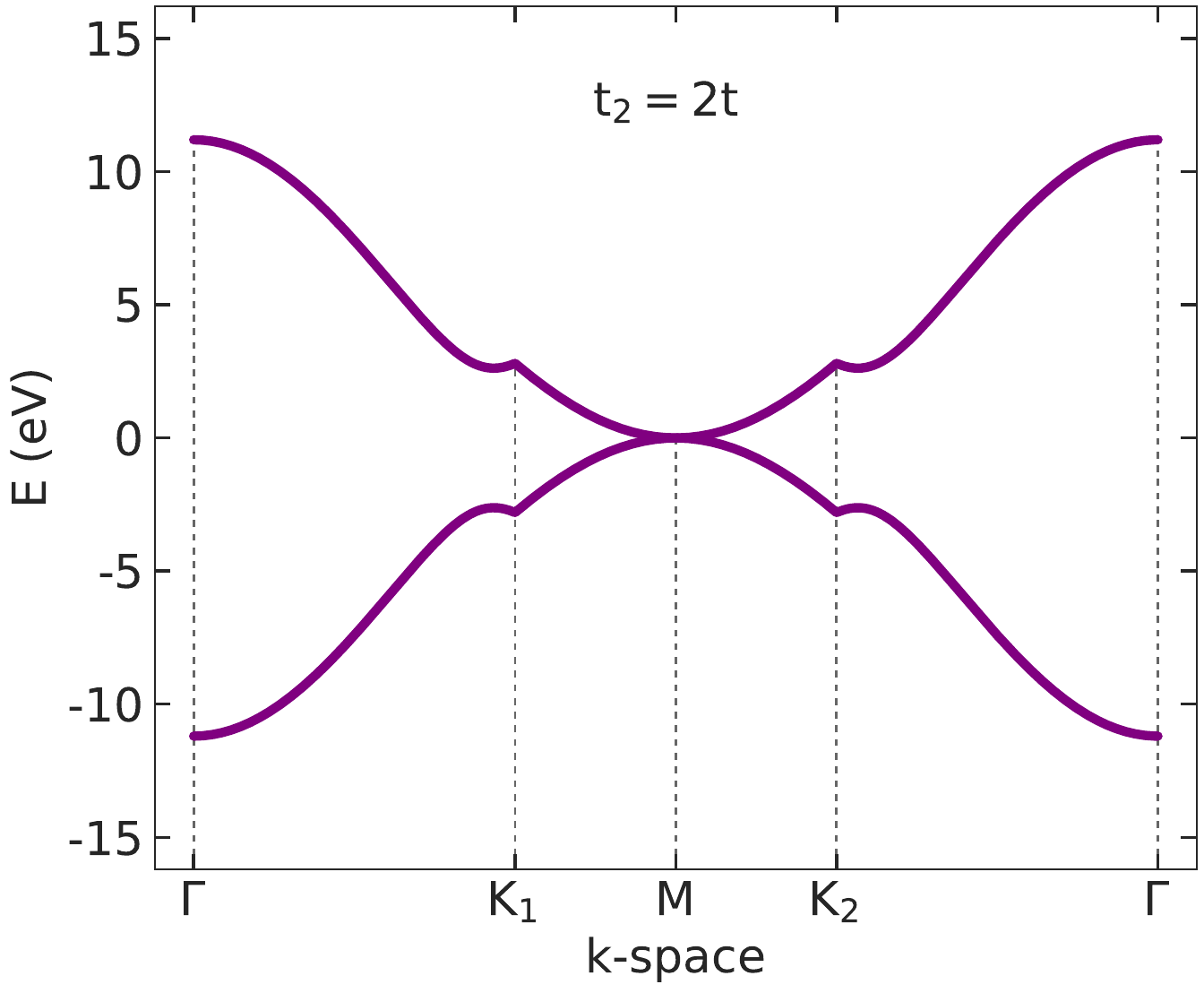}\label{fig:2d}}
\caption{(Color online) (a) Anisotropic energy band dispersion of a semi-Dirac system is shown. The dispersion is linear along the $y$-direction and quadratic along the $x$-direction. Here $a$ is set to be unity. (b) Brillouin zone with different high-symmetry points is shown for Dirac. (c) and (d) The dispersion along the high symmetry points $\Gamma$ $\rightarrow$ $K_1$ $\rightarrow$ $M$ $\rightarrow$ $K_2$ $\rightarrow$ $\Gamma$ for different strength of hopping parameters $t_2=t$ (Dirac) and $t_2=2t$ (semi-Dirac) respectively. Here we put $t=2.8$eV.}
\label{fig:2}
\end{center}
\end{figure*}

We study the tight-binding model on the honeycomb lattice with anisotropic hopping that leads to semi-Dirac electronic spectra at low energy. 
More precisely, the hopping energy to one of the neighbours ($t_2$) is different than the other two ($t$) as shown in Fig.~\ref{fig:1}. 
It is also instructive to look at the full dispersion with the following three nearest neighbor vectors in real space,
$\vec{\delta_1} = \big(0, a\big)$;
$\vec{\delta_2} = \Big(\frac{\sqrt{3}a}{2}, -\frac{a}{2}\Big)$ and
$\vec{\delta_3} = \Big(-\frac{\sqrt{3}a}{2}, -\frac{a}{2}\Big)$, where $a$ is the lattice constant.
\begin{widetext}
The dispersion relation for a semi-Dirac system can be written as,
%\begin{tiny}
\begin{align}
E(k)= 
\pm \sqrt{2t^2 + t_2^2 +2t^2\cos \sqrt{3}k_xa + 4tt_2 \cos(3k_ya/2) \cos(\sqrt3k_xa/2)}.
\label{eq.3}
\end{align}
%\end{tiny}
\end{widetext}
The above expression in Eq.~(\ref{eq.3}) is plotted in Fig.~\ref{fig:2a}. The Brillouin zone with high-symmetry points for $t_2=t$ is shown in Fig.~\ref{fig:2b}. For the Dirac case (that is, $t_2=t$), the dispersion shows that the Dirac points touch at the $K_1$ and $K_2$ points at the Brillouin zone corners as shown in Fig.~\ref{fig:2c}. With increasing the strength of the parameter $t_{2}$, the two Dirac points originally located at $K_1$ ($\frac{2\pi}{3a}$,$\frac{2\pi}{\sqrt{3}a}$) and $K_2$ ($-\frac{2\pi}{3a}$,$\frac{2\pi}{\sqrt{3}a}$) move closer till they merge at the $M$ point resulting in a semi-Dirac spectrum (see Fig.~\ref{fig:2d}). As mentioned earlier, such manipulation of the Dirac points and their eventual merger have been achieved in honeycomb optical lattices\cite{tarruell}. 
Thus by fixing $t_{2} = 2t$ and focusing on
the $M$ point (0,$\frac{2\pi}{\sqrt{3}a}$), the low-energy effective Hamiltonian based on the tight-binding model for a semi-Dirac system, apart from a constant term, can be written as\cite{kush,chen,firoz},
\begin{equation}
{\it{H_0}}=\frac{p^2_x \sigma_x}{2m^*} + v_F p_y \sigma_y
\label{eq.1}
\end{equation}
where $p_x$ and $p_y$ are the momenta along the $x$ and the $y$ directions respectively. $\sigma_x$ and $\sigma_y$ are the Pauli spin matrices in the pseudospin space. The velocity along the $p_y$ direction, $v_F$, and the effective mass, $m^*$ corresponding to the parabolic dispersion along $p_x$ are expressed as $v_F= {3ta}/{\hbar}$ and $m^*=2\hbar/3ta^2$. Henceforth we set $a=1$. The dispersion relation corresponding to Eq.~(\ref{eq.1}) ignoring a constant shift in energy can be written as,
\begin{equation}
E=\pm \sqrt{(\hbar v_{F}k_{y})^2+\bigg(\frac{\hbar^2k_{x}^2}{2m^*}\bigg)^2} 
\label{eq.2}
\end{equation}
where `+' denotes for the conduction band and `-' stands for the valence band. Equation (\ref{eq.2}) shows that the dispersion is linear (Dirac-like) along $y$-direction, whereas the dispersion along the $x$-direction is quadratic (non-relativistic), the combination of which results in the semi-Dirac dispersion. The three-dimensional plot in Fig.~\ref{fig:2a} indicates the anisotropic band structure in a semi-Dirac system.\par

\section{The Landau Levels}\label{III}
To include a magnetic field, we shall work with a semi-Dirac nanoribbon which is infinitely long along $x$, but has a finite width along $y$. We apply a uniform magnetic field, $\mathbf{B}=B\hat{z}$ perpendicular to the plane of the ribbon. Owing to the presence of the vector potential $\vec{A}$, each tight-binding wave-function picks up an extra phase term. We have chosen the Landau gauge as $\vec{A} = (-By, 0, 0)$ such that the translational invariance along the $x$-direction remains unaltered under the choice of the gauge. Hence, the momentum along the $x$-direction is conserved and acts as a good quantum number. To make $k_x$ a dimensionless quantity, we have absorbed the lattice spacing $a$ into the definition of $k_x$. The ribbon width is such that it has $N$ unit cells along the $y$-axis (where the index $n$ for the unit cells takes $\in 0$.....$N-1$) as shown in Fig.~\ref{fig:3}.  
The tight-binding Hamiltonian in the presence of magnetic field has the form,
\begin{align}
 H= -\sum_{\langle{ij}\rangle} (t_{ij}a^{\dagger}_{i} b_{j} +h.c.)
 \label{ham}
\end{align}
where $a^{\dagger}_{i}$ ($b_{j}$) creates (annihilates) an electron on sublattice $A$ ($B$). $t_{ij}$ is the hopping amplitude between nearest neighbor sites, which obtain a phase due to the magnetic field by the Peierls substitution, namely, $ t_{ij} = t \rightarrow te^{2i\pi\phi_{ij}}$ (here $t$ denotes both $t$ or $t_2$). $\phi_{ij}$ is the magnetic flux and is given by the line integral of the vector potential from a site $i$ to a site $j$, namely, $\phi_{ij}=e/h\int_{i}^{j} \vec{A}.d\vec{l}$. The flux is usually denoted in terms of the flux quantum $\phi_0=h/e$ ($h$ is Planck$'$s constant and $e$ is the magnitude of the electron charge).
\begin{figure}[h]
\begin{center}
\subfloat{\includegraphics[width=0.45\textwidth]{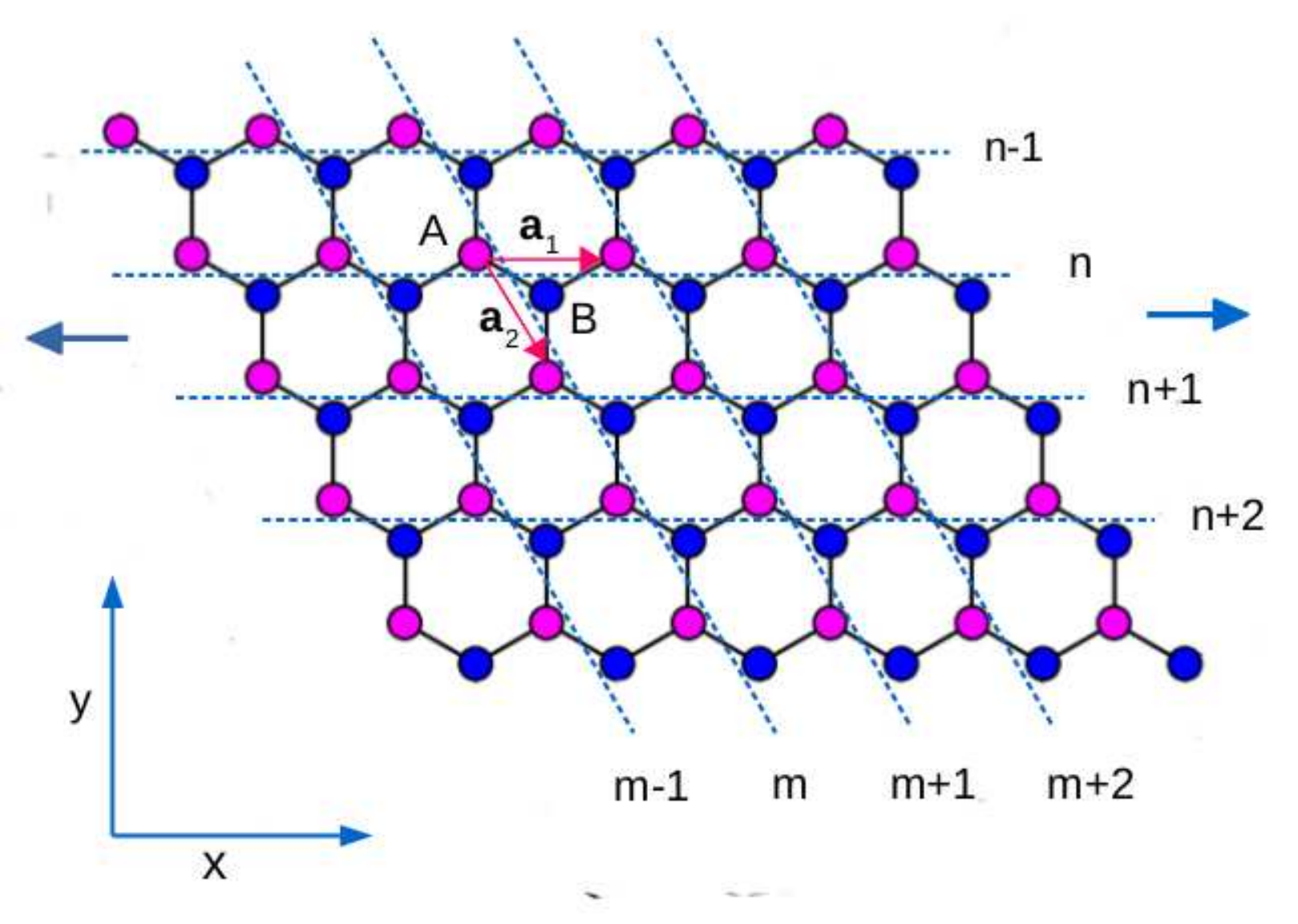}\label{fig:3}}
\caption{(Color online) Zigzag nanoribbon of a honeycomb lattice is shown. The magenta and blue circles represent the $A$ and $B$ sublattices respectively. ${\vec {a}}_{1}$ and ${\vec{a}}_2$ are the primitive vectors. $(m,n)$ labels the positions of the unit cells along $x$ and $y$ directions. The ribbon is infinite along the $x$-direction shown by the arrow on both side.}
\label{fig:3}
\end{center}
\end{figure}
Thus the tight-binding Hamiltonian in presence of the perpendicular magnetic field can be written in terms of $m$ and $n$ (where $m$ increases along the $x$-direction and $n$ increases along the negative $y$-direction) (see Fig.~\ref{fig:3})\cite{castro},
\begin{align} 
\mathcal{H} = &
 - \sum\limits_{\langle{mn}\rangle} \Big[t e^{i\pi(\phi/\phi_0)n[(1+\alpha)/2]}a^{\dagger}(m,n)b(m,n) \nonumber
\\
&
+ t e^{-i\pi(\phi/\phi_0)n} a^{\dagger}(m,n)b(m-1,n-(1-\alpha)/2)\nonumber 
\\
&
+ t_2~e^{i\pi(\phi/\phi_0)n[(\alpha-1)/2]} a^{\dagger}(m,n)b(m,n-\alpha)+h.c.\Big]
\label{h1}
\end{align}
where the summation $\langle{mn}\rangle$ is over the nearest neighbors. $a^{\dagger}(m,n)$ and $b(m,n)$ denote the creation and annihilation operators at the ($m$,$n$) site, respectively. 
Equation ({\ref{h1}}) for a zigzag semi-Dirac ribbon ($\alpha=1$) reduces to 
\begin{align}
\mathcal{H} = &
- \sum\limits_{\langle{mn}\rangle} \Big[t e^{i\pi(\phi/\phi_0)n}a^{\dagger}(m,n)
b(m,n)+ t e^{-i\pi(\phi/\phi_0)n} \nonumber
\\
&
a^{\dagger}(m,n)b(m-1,n)  
+ t_2~a^{\dagger}(m,n)b(m,n-1)+h.c.\Big]
\label{h2}
\end{align}
\begin{figure}[h]
\begin{center}
\subfloat[]{\includegraphics[width=0.24\textwidth]{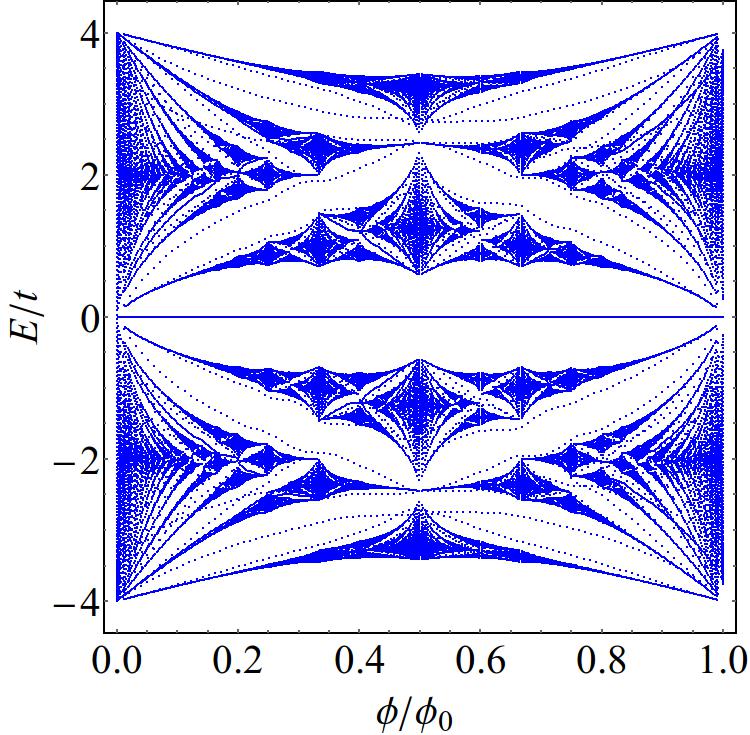}\label{fig:4a}} \hspace*{0.01 cm}
\subfloat[]{\includegraphics[width=0.24\textwidth]{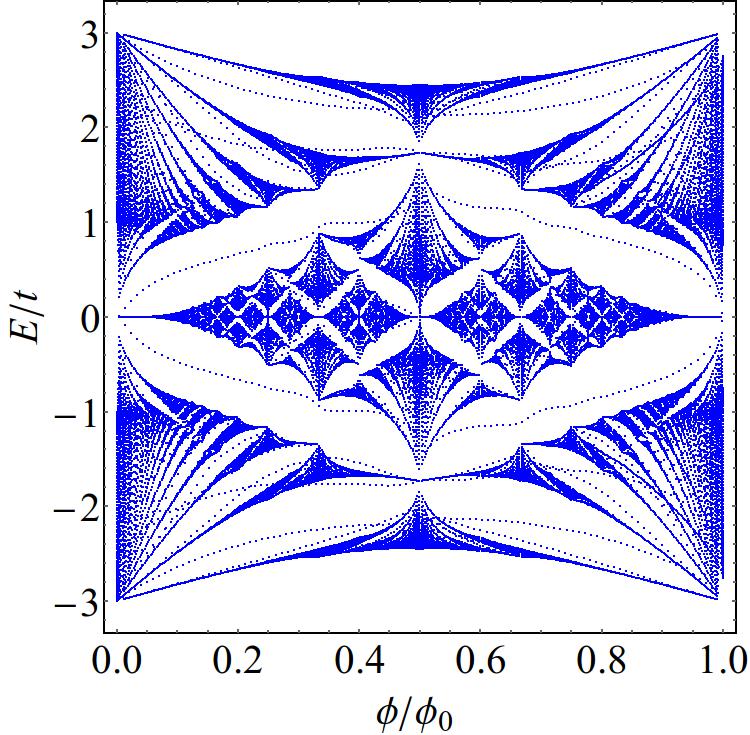}\label{fig:4b}}
\caption{(Color online) Hofstadter butterfly spectrum is plotted for as a function of $\phi/\phi_0$ for (a) $t_2=2t$ (semi-Dirac) and (b) $t_2=t$ (Dirac).}
\label{fig:4}
\end{center}
\end{figure}
Using the above Hamiltonian as mentioned in Eq.~(\ref{h2}) we have numerically calculated the Hofstadter butterfly\cite{rammal} as well as the Landau level spectrum for the number of unit cells $N=100$. Figure~\ref{fig:4a} shows fractal spectra plotted as a function of magnetic flux, $\phi/\phi_{0}$ for a semi-Dirac nanoribbon. It can be seen clearly that there occurs opening of a central gap with a flat band at zero energy. The gap gets larger along with the two identical spectra that emerge from the conduction and the valence bands by tuning $t_2$. For comparison, the same is plotted for the Dirac system ($t_2=t$) as shown in Fig.~\ref{fig:4b}. We can see that there is no gap at zero energy with the flat band when one goes from $t_2=2t$ to $t_2=t$. 
\begin{figure*}[!ht]
\centering
\subfloat[]{\includegraphics[width=0.243\textwidth]{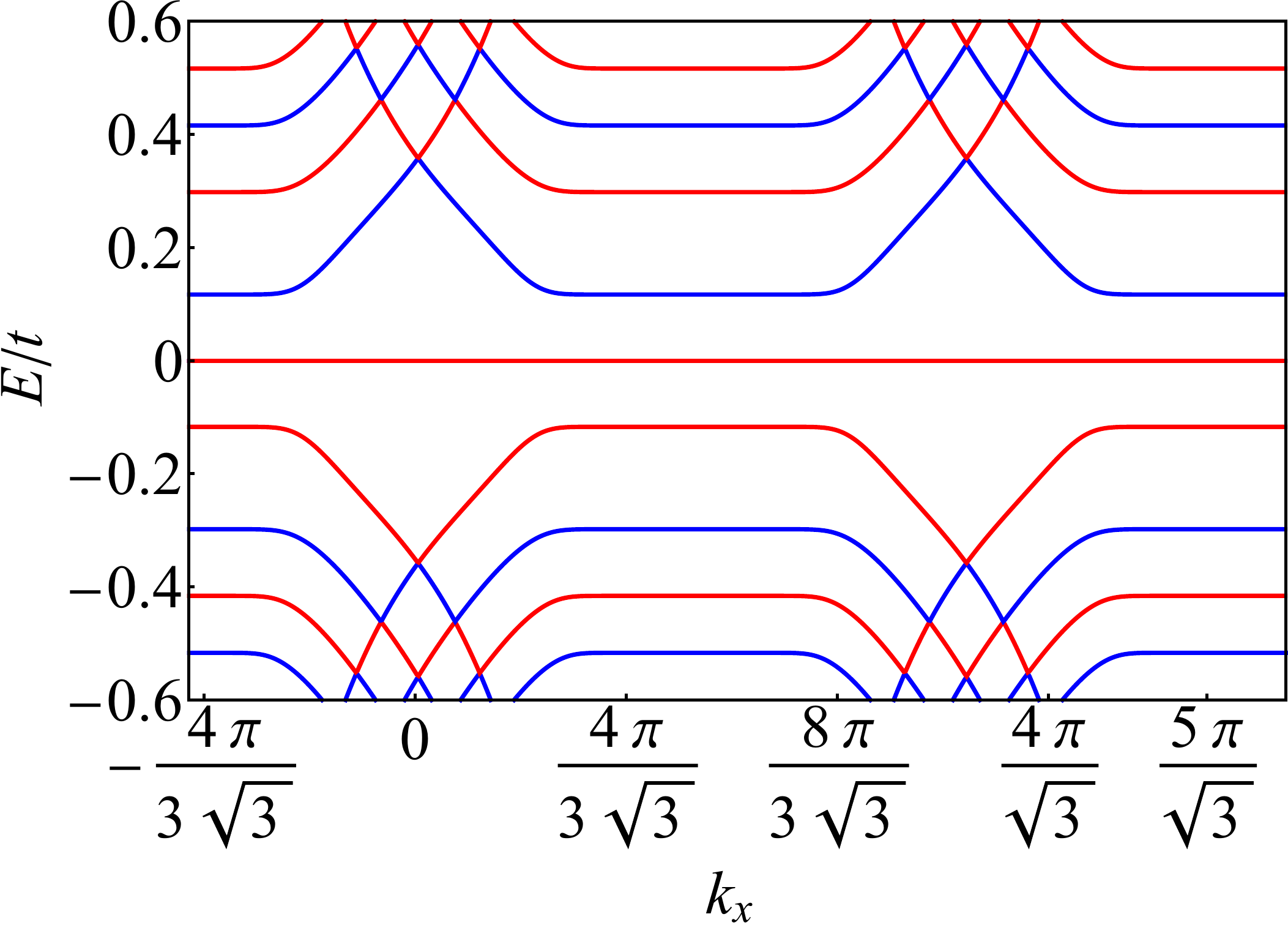}\label{fig:5a}} \hspace{0.02 cm}
\subfloat[]{\includegraphics[width=0.243\textwidth]{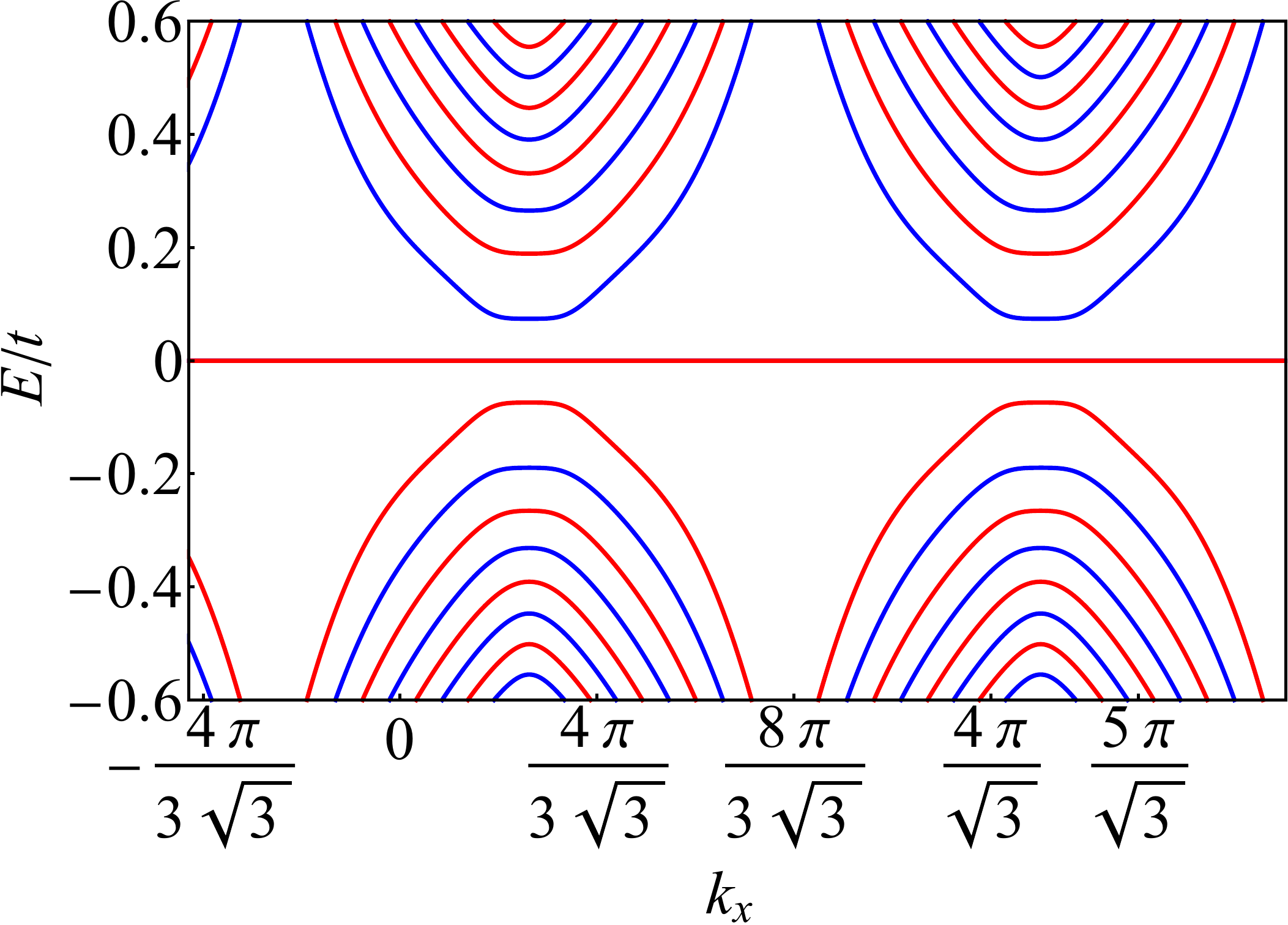}\label{fig:5b}} \hspace{0.02 cm}
\subfloat[]{\includegraphics[width=0.243\textwidth]{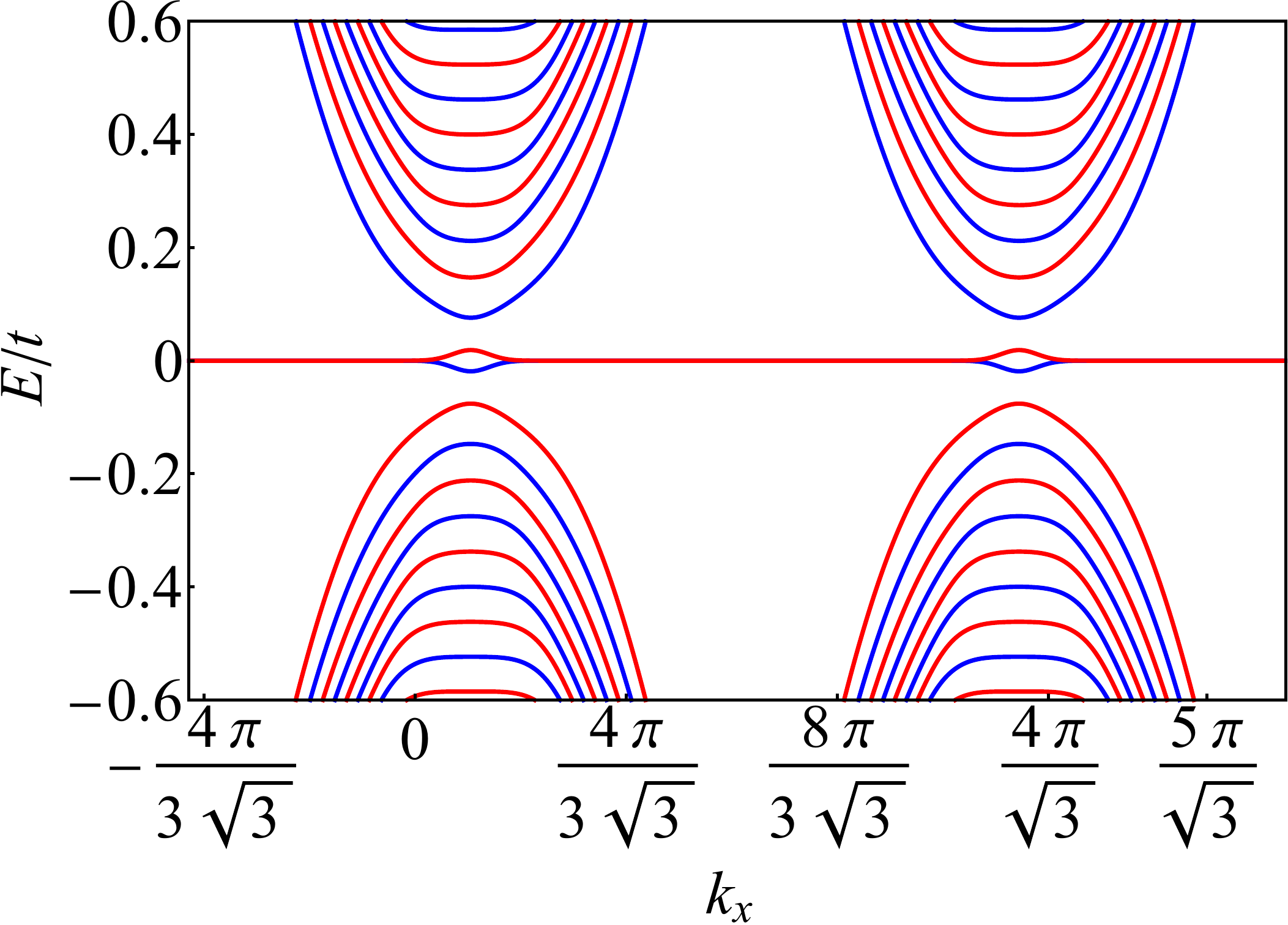}\label{fig:5c}} \hspace{0.02 cm}
\subfloat[]{\includegraphics[width=0.243\textwidth]{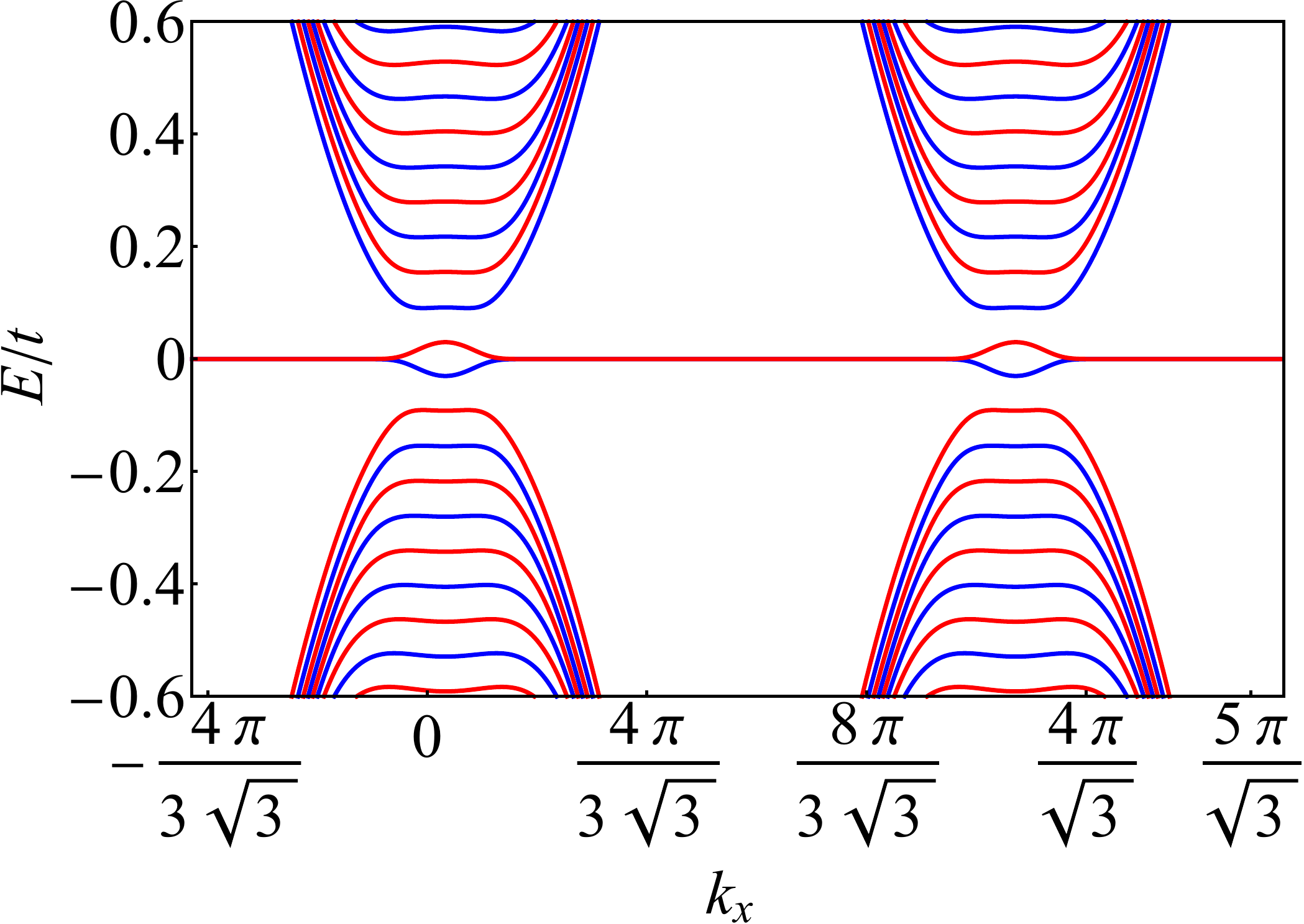}\label{fig:5d}}\\
\subfloat[]{\includegraphics[width=0.243\textwidth]{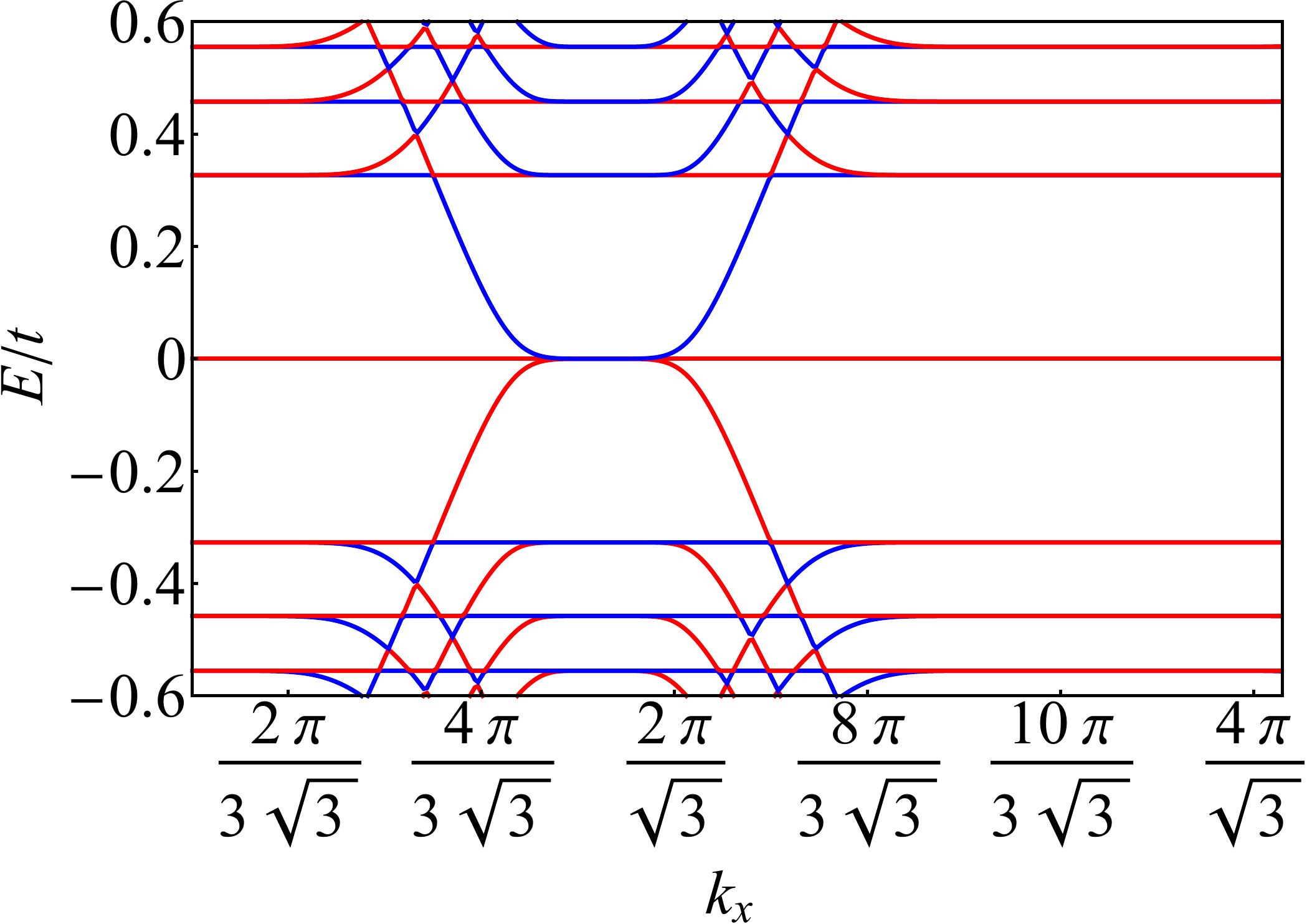}\label{fig:5e}} \hspace{0.02 cm}
\subfloat[]{\includegraphics[width=0.243\textwidth]{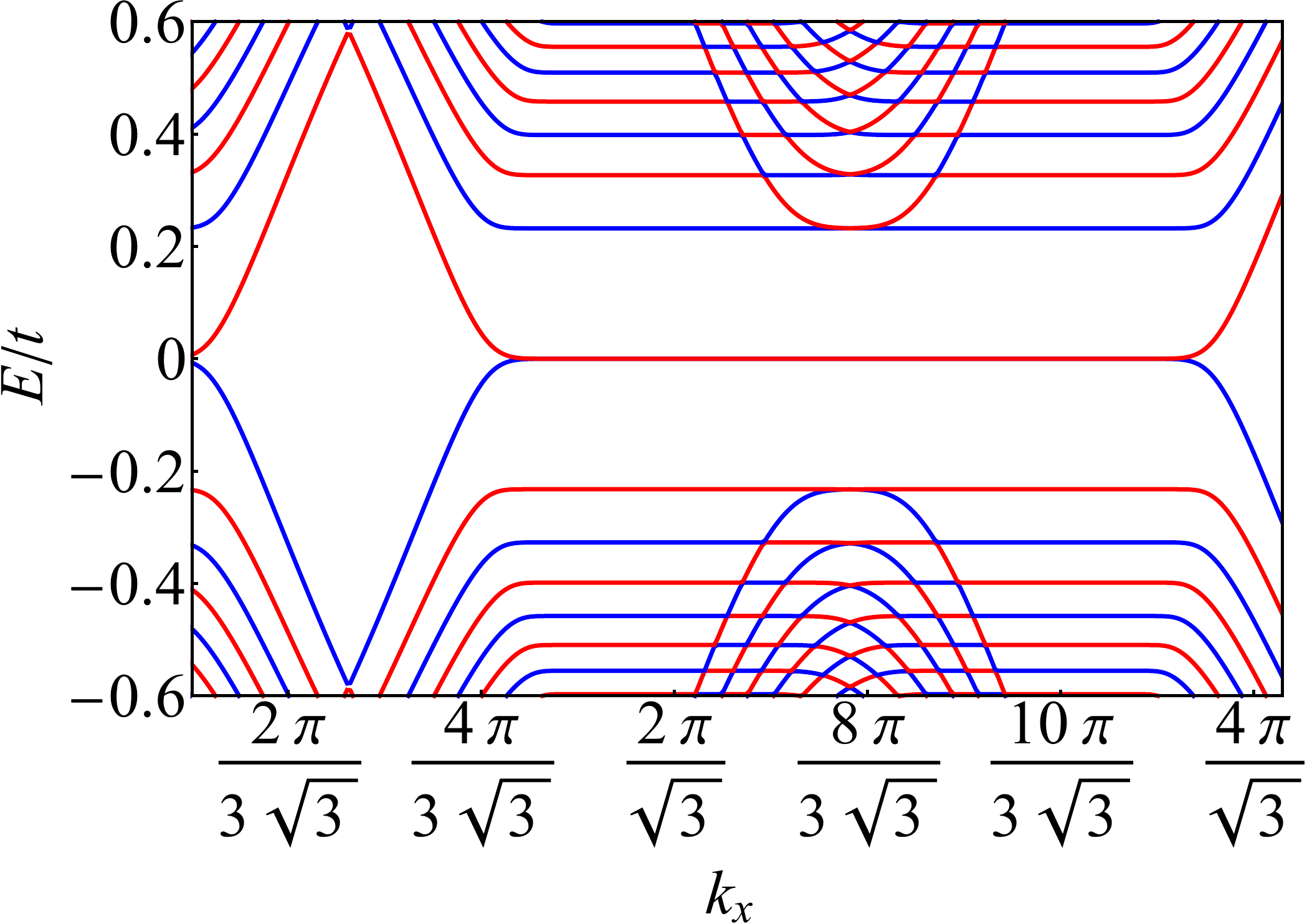}\label{fig:5f}} \hspace{0.02 cm}
\subfloat[]{\includegraphics[width=0.243\textwidth]{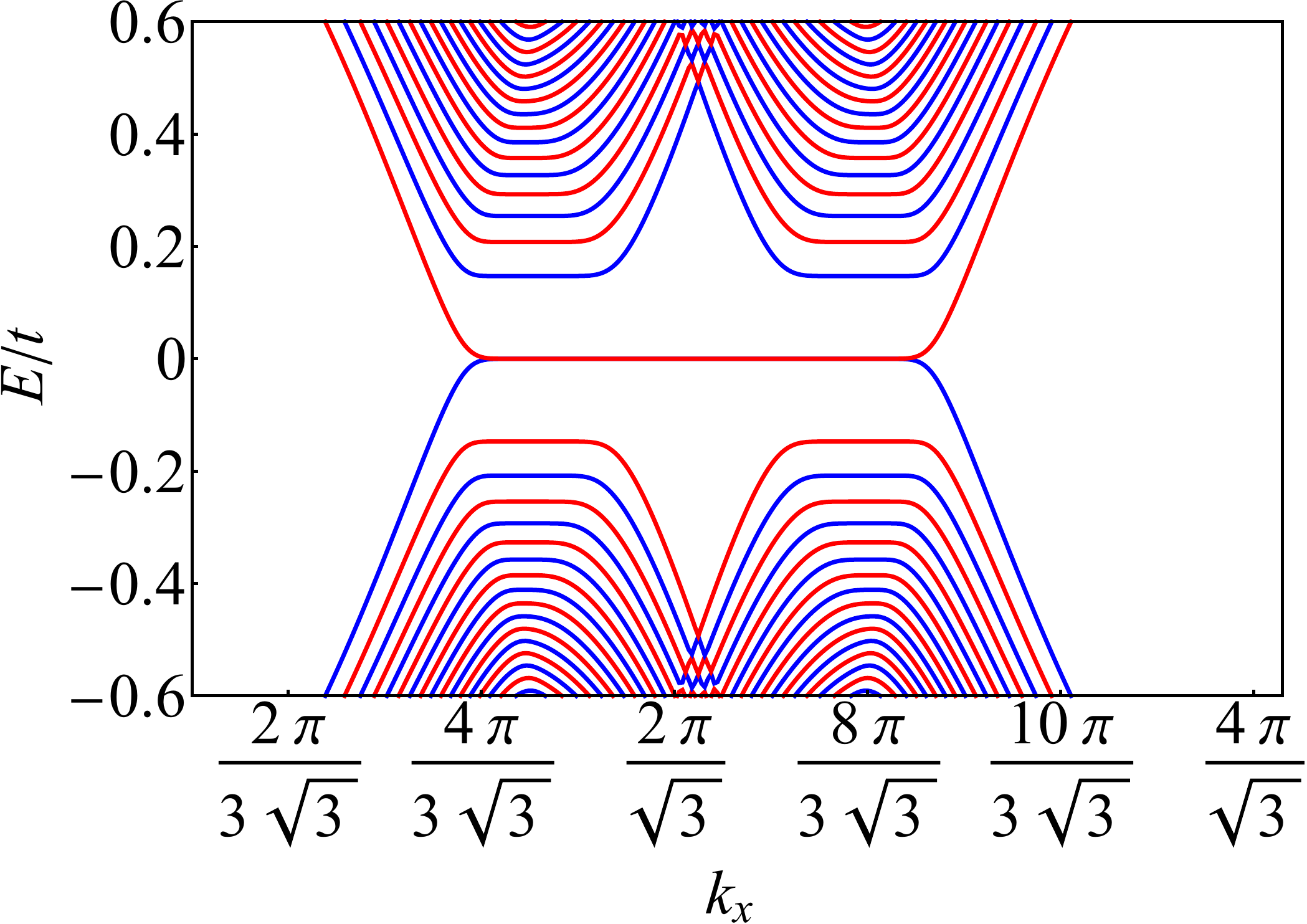}\label{fig:5g}} \hspace{0.02 cm}
\subfloat[]{\includegraphics[width=0.243\textwidth]{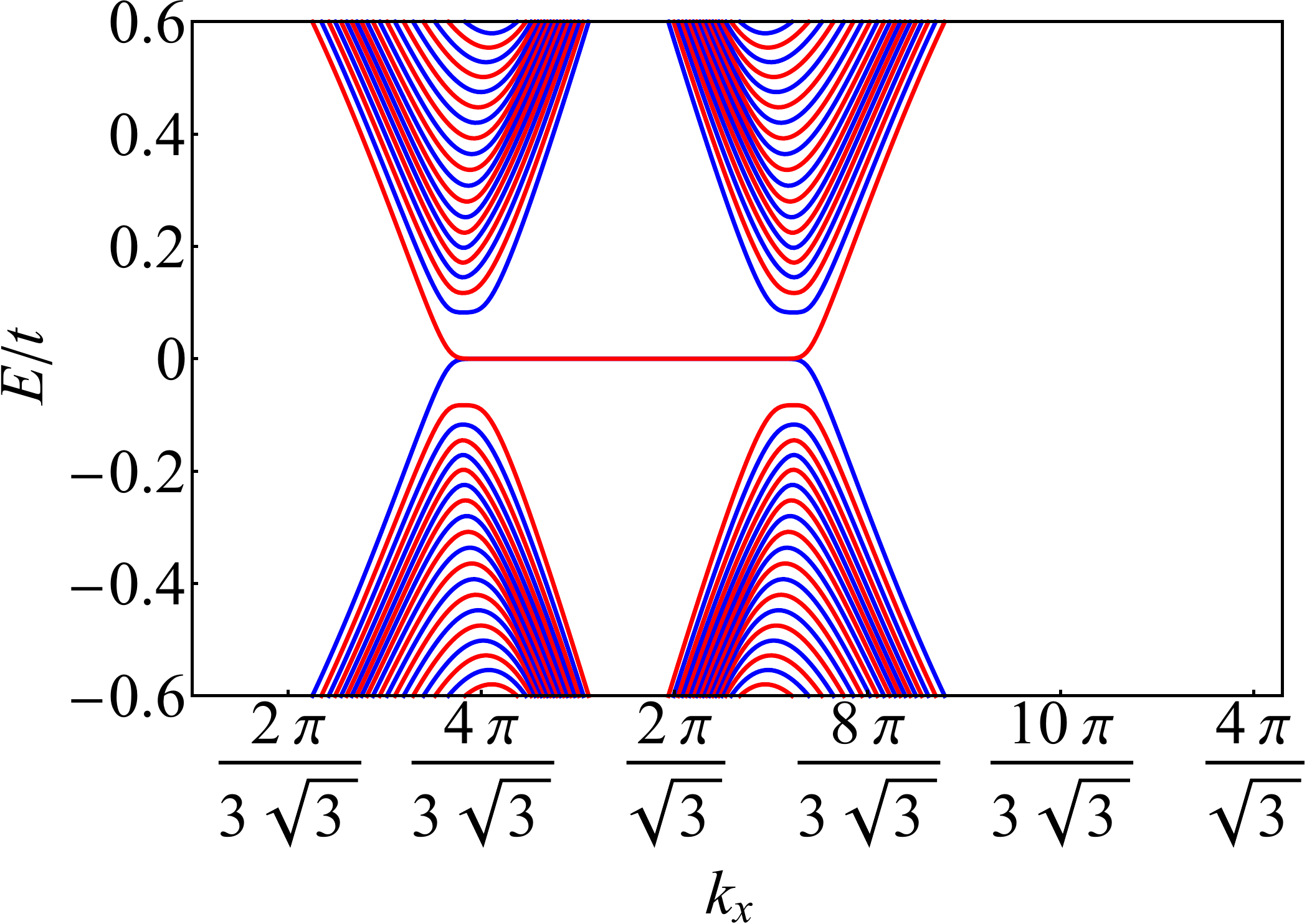}\label{fig:5h}} 
\caption{(Color online) Evolution of Landau levels for a finite strip with $N=100$ unit cells in the presence of a magnetic flux $\phi=\frac{\phi_0}{100}$, $\frac{\phi_0}{200}$,$\frac{\phi_0}{500}$ and $\frac{\phi_0}{1600}$ for (a-d) $t_2=2t$ (semi-Dirac) and (e-h) $t_2=t$ (Dirac).}
\label{fig:5}
\end{figure*}
Figure~\ref{fig:5} shows the Landau level spectra for different values of the magnetic flux (such as $\phi/\phi_{0} = 1/100, 1/200, 1/500, 1/1600$) for the semi-Dirac and the Dirac systems. Figures~\ref{fig:5a}-\ref{fig:5d} show the evolution of the energy levels in presence of the magnetic field for a semi-Dirac nanoribbon ($t_2=2t$). For comparison, we have also plotted the Landau level spectrum for the Dirac case ($t_2=t$) using the same values of the magnetic flux as shown in Fig.~\ref{fig:5e}-\ref{fig:5h}. It is to be noted that in a semi-Dirac system, there is no zero energy bulk state, which implies that the zero energy state in Fig.~\ref{fig:5} is an edge state. On the other hand, zero energy bulk states exist in a Dirac system.
Further, for $t_2=2t$, the Landau levels are not equidistant, since their energies vary as $(n+\frac{1}{2})^{2/3}$ ($n$ being the Landau level index)\cite{pickett} which lies intermediate to the behavior of the Dirac system and the conventional 2DEG. As a consequence, the gap between the Landau levels shrinks as one considers larger $n$. In the case of a Dirac system, since the energies of the Landau levels go as $\sqrt{n}$, its non-equidistant Landau spectra can have a different quantitative behavior from a semi-Dirac system. For a large value of the flux, $\phi$ such as $\phi=\phi_0/100$, the flatness of the energy bands are observed for both the semi-Dirac and the Dirac systems owing to shrinking of the cyclotron radius (see Fig.~\ref{fig:5a} and \ref{fig:5e}). The energy bands become parabolic for $\phi=\phi_0/200$ as seen from Fig.~\ref{fig:5b}. The flatness of the Landau spectrum in the bulk completely vanishes in the semi-Dirac system as compared to a Dirac one (see Fig.~\ref{fig:5b} and \ref{fig:5f}). With lower values of $\phi/\phi_0$, the Landau levels demonstrate a dispersive behavior and start getting flatter for large values of $n$ for $t_2=2t$ (see Fig.~\ref{fig:5c}). In the case of a Dirac system ($t_2=t$), the Landau levels show quite a distinct behavior, where the flat bands become dispersive in the bulk corresponding to larger values of $n$ and lower values of $\phi/\phi_0$ (see Fig.~\ref{fig:5g}). For a small value of the magnetic field, such that the flux is given by, $\phi=\phi_0/1600$, the energy bands eventually become flat in the bulk for the semi-Dirac case as shown in Fig.~\ref{fig:5d}. This is not the case for a Dirac system (see Fig.~\ref{fig:5h}). For all values of $\phi/\phi_0$, the zero-energy mode is completely separated from the bulk bands for a semi-Dirac system as compared to the Dirac case (see any of Fig.~\ref{fig:5}). 
\section{Transport Properties}\label{IV}
To study the transport properties in the presence of a perpendicular magnetic field, we consider a large sample of a lattice model of the semi-Dirac system consisting of millions of atoms. The contribution in transport comes from both the off-diagonal and the diagonal terms as appear in the Kubo formula\cite{kubo1}. The former contributes to the Hall conductivity ($\sigma_{xy}$), whereas the latter leads to individual longitudinal conductivity in different directions ($\sigma_{xx}$ and $\sigma_{yy}$).   
\subsection{Methodology}
In this section, we shall describe the numerical approach, developed by Garcia and his co-workers \cite {rappoprt} which is based on a real-space implementation of the Kubo formalism, where both the diagonal and the off-diagonal conductivities are treated on the same footing. It is known that in the momentum space, the Hall conductivity can be easily obtained in terms of the Berry curvature associated with the bands \cite{kohmoto}. The Kubo formalism can be implemented in real space for obtaining the Hall conductivity\cite{rappoprt} which uses Chebyshev expansions to compute the conductivities. The components of the dc conductivity tensor ($\omega \rightarrow 0$ limit of the ac conductivity) for the non-interacting electrons are given by the Kubo-Bastin formula\cite{kubo1, kubo2} which can be written as\cite{bastin,rappoprt,ortmann},
\begin{align}
\sigma_{\alpha\beta} (\mu, T) = & \frac{i e^2 \hbar}{A}\int_{-\infty}^{\infty} d\varepsilon f(\varepsilon) Tr \Big<v_{\alpha} \delta(\varepsilon-H)v_\beta \frac{dG^{+}(\varepsilon)}{d\varepsilon} \nonumber
\\
&
-v_{\alpha} \frac{dG^{-}(\varepsilon)}{d\varepsilon}v_\beta\delta(\varepsilon-H)\Big>
 \label{eq.7}
\end{align} 
where $T$ is the temperature, $\mu$ is the chemical potential, $v_\alpha$ is the $\alpha$ component of the velocity operator, $A$ is the area of the sample, $f(\varepsilon)$ is the Fermi-Dirac distribution and $G^{\pm}(\varepsilon, H)= \frac{1}{\varepsilon-H\pm i\eta}$ are the advanced (+) and retarded (-) Green's functions. Using the Kernel Polynomial Method (KPM)\cite{weisse}, the rescaled delta and Green's function can be expanded in terms of the Chebyshev polynomials, whence Eq.~(\ref{eq.7}) becomes,
\begin{equation}
\sigma_{\alpha\beta} (\mu, T) = \frac{4 e^2 \hbar}{\pi A} \frac{4}{(\Delta E)^2} \int_{-1}^{1} d{\tilde{\varepsilon}} \frac{f(\tilde{\varepsilon})}{(1-{\tilde{\varepsilon}^2})^2} \sum_{m,n} \Gamma_{nm} (\tilde{\varepsilon}) \mu_{nm}^{\alpha\beta} (\tilde{H})
\label{eq.8}
\end{equation} 
where $\Delta E$ is the range of the energy spectrum, $\tilde{\varepsilon}$ is the rescaled energy whose upper and lower bounds are $+1$ and $-1$ respectively and $\tilde{H}$ is the rescaled Hamiltonian. $\Gamma_{nm} (\tilde{\varepsilon})$ and $\mu_{nm}^{\alpha\beta} (\tilde{H})$ are the functions of rescaled energy and Hamiltonian respectively. The energy dependent scalar function, $\Gamma_{nm} (\tilde{\varepsilon})$ can be written as, 
\begin{align}
 \Gamma_{nm} (\tilde{\varepsilon})\equiv &(\tilde{\varepsilon}-in\sqrt{1-\tilde{\varepsilon}^2})e^{in~arccos(\tilde{\varepsilon})} T_m(\tilde{\varepsilon}) \nonumber
 \\
 &
+ (\tilde{\varepsilon}+im\sqrt{1-\tilde{\varepsilon}^2})e^{-im~arccos(\tilde{\varepsilon})} T_n(\tilde{\varepsilon}) 
\end{align}
and the Hamiltonian-dependent term which involves products of polynomial expansions can be written as,
\begin{equation}
\mu_{nm}^{\alpha\beta}(\tilde{H})= \frac{g_m g_n}{(1+\delta_{n0})(1+\delta_{m0})} Tr[v_{\alpha}T_m(\tilde{H})v_{\beta}T_n(\tilde{H})]
\end{equation}
where the Chebyshev polynomials, $T_m(x)$ obey the recurrence relation,
\begin{equation}
T_m(x)= 2x T_{m-1}(x)-T_{m-2}(x)
\end{equation}
The Jackson kernel, $g_m$ is used to smoothen out the Gibbs oscillations which arise due to the truncation of the expansion in Eq.~(\ref{eq.8}) \cite{weisse,silvera}.\par
\begin{figure}[ht]
\begin{center}
\subfloat[]{\includegraphics[width=0.25\textwidth]{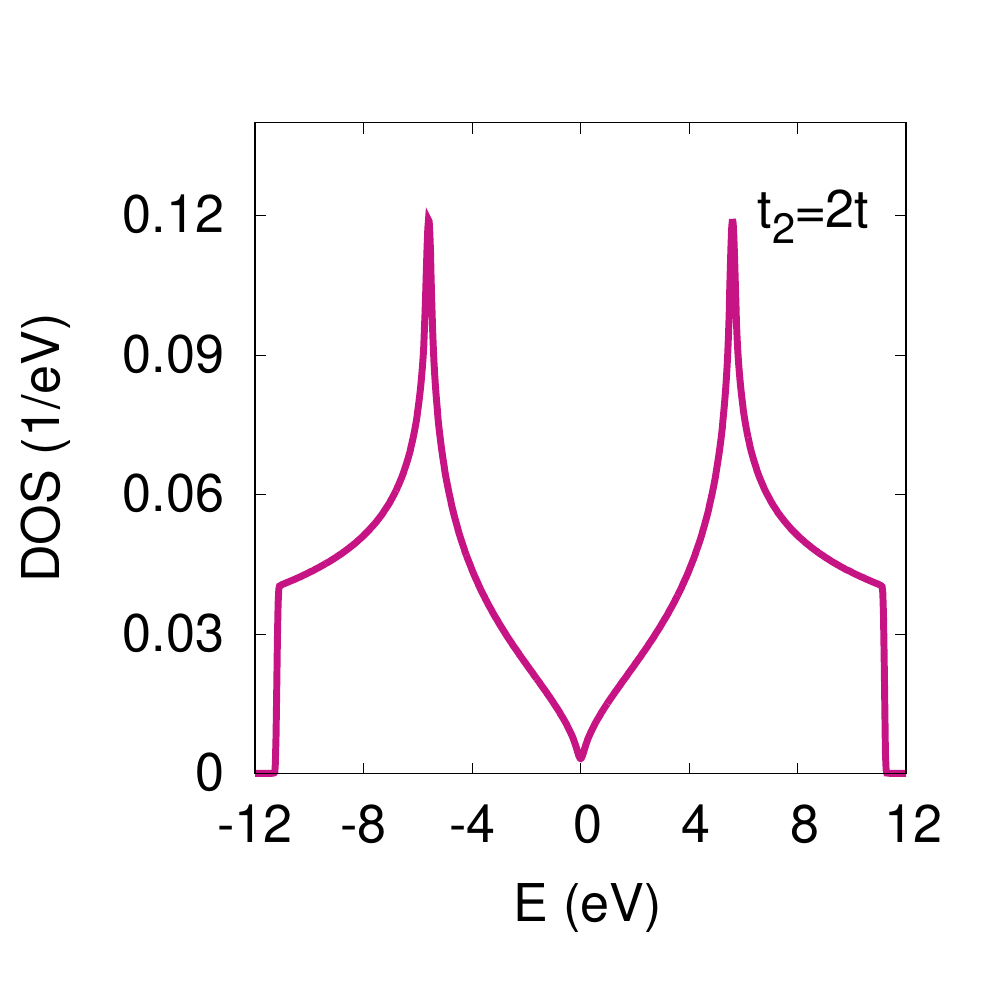}\label{fig:6a}} \hspace*{0.01 cm}
\subfloat[]{\includegraphics[width=0.25\textwidth]{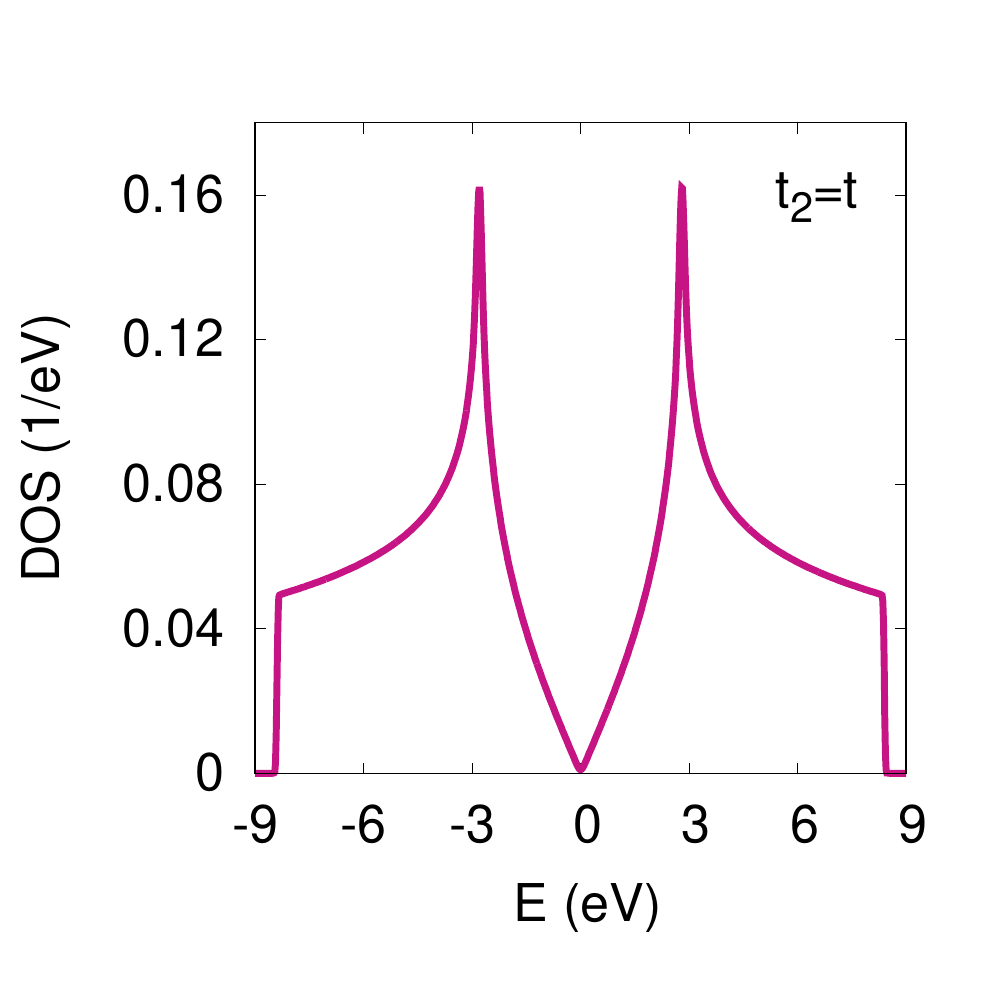}\label{fig:6b}}       
\caption{(Color online) Density of states (in units of 1/eV) is plotted as a function of energy, $E$ (in units of eV) for (a) $t_2=2t$ (semi-Dirac) and (b) $t_2=t$ (Dirac). We put $t=2.8$eV in the calculation.}
\label{fig:6}
\end{center}
\end{figure}
\begin{figure*}[!ht]
\centering
\subfloat[]{\includegraphics[width=0.35\textwidth]{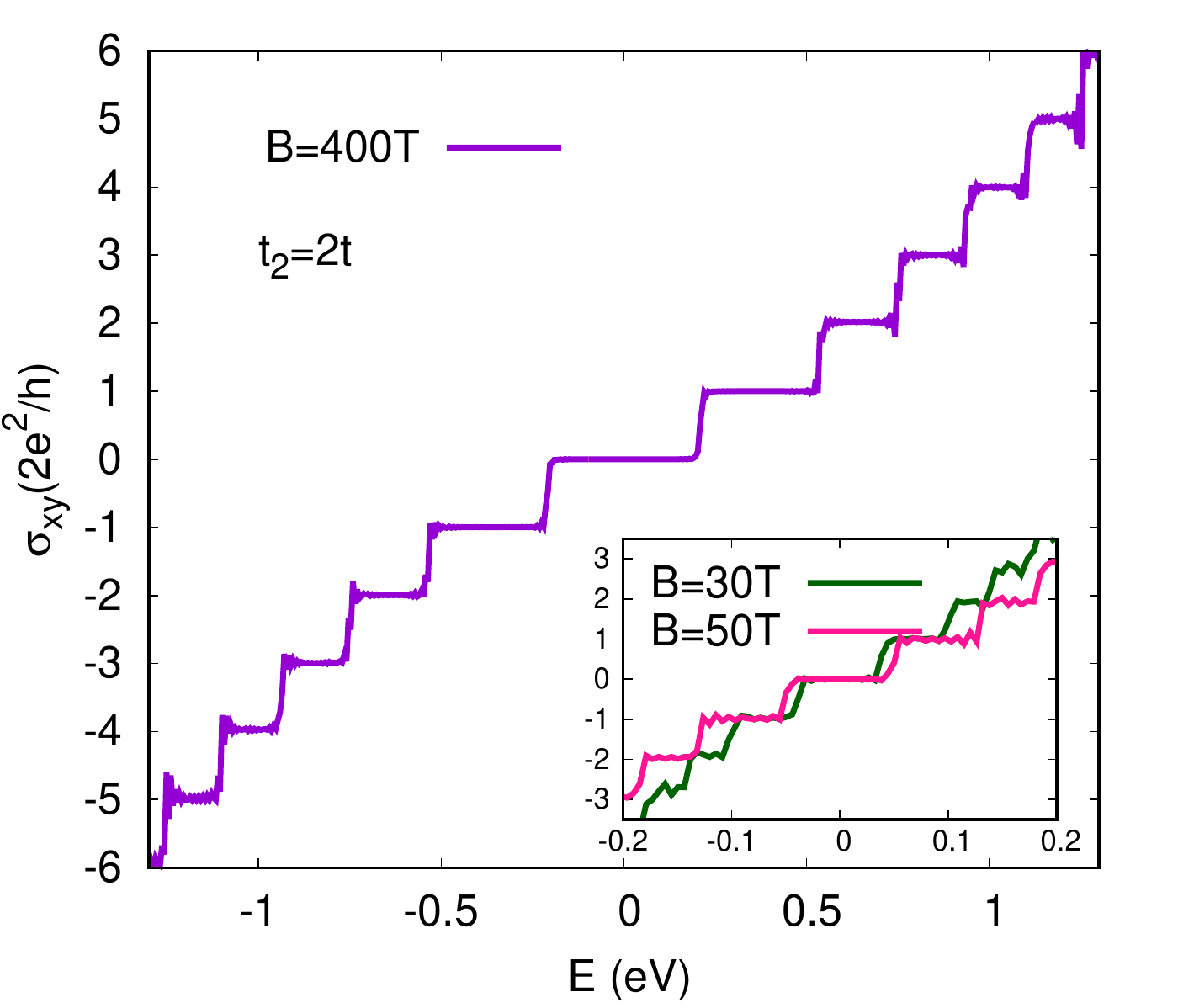}\label{fig:7a}} \hspace*{ 2.5 cm}
\subfloat[]{\includegraphics[width=0.35\textwidth]{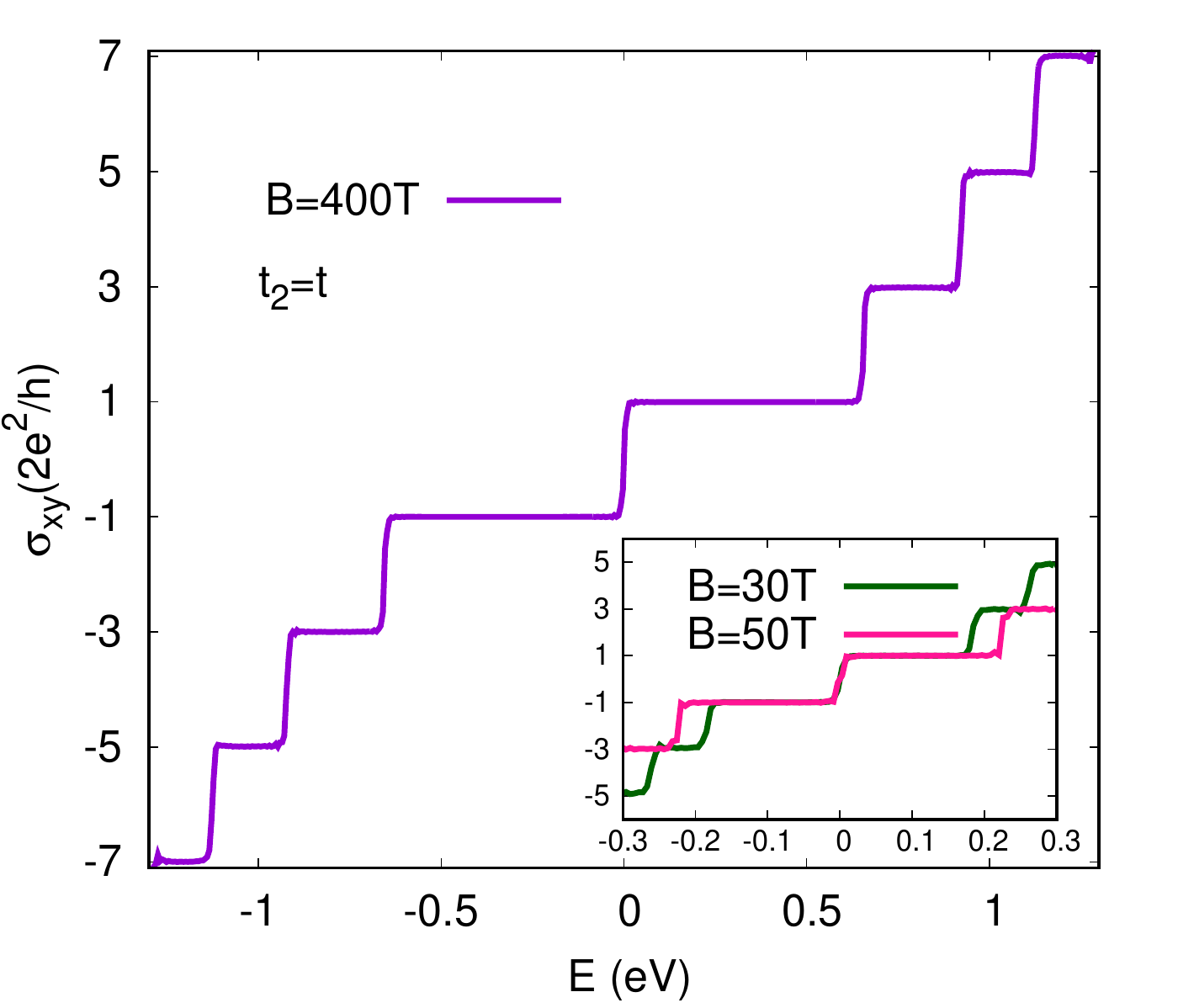}\label{fig:7b}}\\
\caption{(Color online) Hall conductivity, $\sigma_{xy}$ (in units of $2e^2/h$) is plotted as a function of  Fermi energy, $E$ (in units of eV) for (a) $t_2$ = $2t$ (semi-Dirac) and (b) $t_2$ = $t$ (Dirac) for a very high field (400T) and a moderate field (30T and 50T). Here $t$ is taken as 2.8eV.}
\label{fig:7}
\end{figure*} 
The density of states (DOS) can be calculated using an efficient algorithm based on the evolution of the time-dependent Schr$\ddot{o}$dinger equation. We use a random superposition of all basis states as an initial state $|\phi(0)\rangle$,
\begin{equation}
|\phi(0)\rangle=\sum_{i}a_i|i\rangle
\end{equation}
where $|i\rangle$ denote the basis states and $a_i$ are the normalized random complex numbers. Applying the Fourier transformation to the correlation function, $\langle\phi(0)|e^{-iHt}|\phi(0)\rangle$ we get the DOS as \cite{yuan},
\begin{equation}
DOS=\frac{1}{2\pi}\int_{-\infty}^{\infty}e^{i\epsilon t}\langle\phi(0)|e^{-iHt}|\phi(0)\rangle dt
\end{equation}
where $t$ denotes time.
%Also the density of state (DOS) is given by,
%\begin{equation}
%DOS=\frac{1}{D}\sum_{k=0}^{D-1}\delta(E-E_k)
%\end{equation}
\subsection{Longitudinal and Hall conductivities}
Using the above mentioned efficient numerical approach, we calculate the DOS in the absence and the presence of a magnetic field, the longitudinal conductivity in both $x$ ($\sigma_{xx}$) and $y$ ($\sigma_{yy}$) directions and the Hall conductivity ($\sigma_{xy}$) for the semi-Dirac system ($t_2=2t$). To compare between the Dirac and the semi-Dirac systems, we have also shown results for the Dirac ($t_2=t$) case simultaneously. In our simulation, we consider a lattice of $5120$ unit cells in each of the $x$ and $y$ directions (that is, a sample size denoted by ($L_{x},L_{y}$) to be ($5120, 5120$)). We apply periodic boundary conditions for all our numerical results. We set the nearest neighbor hopping parameter $t=2.8$eV. We adopt a large number of Chebyshev moments, $M$, since the energy resolution of the KPM and the convergence of the peaks of $\sigma_{xx}$ depend on $M$. We have used $M=6144$ here\cite{rappoprt}. The system size and the truncation order can be enhanced to reduce the fluctuations.\par
We have plotted the results for the DOS in the absence of a magnetic field for semi-Dirac and Dirac systems as shown in Fig.~\ref{fig:6}. In the case of semi-Dirac systems, the DOS is proportional to $\sqrt{|E|}$ near zero energy (see Fig.~\ref{fig:6b}), whereas for the Dirac case the DOS near the zero energy varies as $|E|$ as shown in Fig.~\ref{fig:6a}. The energy bounds for the semi-Dirac are $E_{min}=-4t$ and $E_{max}=4t$.\par
Next, we have shown the results for Hall conductivity ($\sigma_{xy}$) for moderate values of the magnetic field as well as extremely high fields in Fig.~\ref{fig:7}. Generally higher values of the magnetic field require smaller system size and hence the fewer number of Chebyshev moments have to be computed. This yields faster convergence of the Hall conductivity in the limit of a large magnetic field. For $t_2=2t$ (semi-Dirac), the Hall conductivity ($\sigma_{xy}$) is plotted as a function of Fermi energy, $E$ for the large value of the field $B=400$T (as shown in the main frame of Fig.~\ref{fig:7a}). To relate this result to the recent experiments \cite{young,dean,hunt} performed for realistic values of magnetic field on a Dirac system, we have also plotted the Hall conductivity for moderate values of the field, namely 30T (green curve) and 50T (pink curve) as shown in the inset of Fig.~\ref{fig:7a}. The quantization of the plateaus is similar to that of a conventional 2DEG with a parabolic band dispersion in a sense that the conductance quantization happens at $\sigma_{xy}=2ne^2/h$ where $n$ takes integer values $0$, $\pm1$, $\pm2$, $\pm3$, $\pm4$... in units of $2e^2/h$. The plot in the inset shows that the plateau step can be obtained with good accuracy even in the case of realistic values of the magnetic field. The difference between the semi-Dirac and the Dirac cases lies in the fact that the fluctuations in the plateau step become prominent with lowering of the field, especially at higher values of the Fermi energy. Further lowering of the magnetic field will reduce the sharpness of the plateaus due to the effect of finite energy resolution and finite size of the sample. These are the artifacts of the method used here. In the Dirac system ($t_2=t$), the well-known Hall quantization at $\sigma_{xy}= 2(2n+1)e^2/h$ is observed for very high magnetic field, namely $B=400$T as shown in the main frame of Fig.~\ref{fig:7b}. The inset shows the same for realistic values of the magnetic field. The Hall conductivity plot ensures that there is a transition from a half-integer to an integer quantum Hall effect as we go from a Dirac to a semi-Dirac system by tuning $t_2$ (Fig.~\ref{fig:7a} and \ref{fig:7b}). The reason can be drawn from the fact that although the band dispersion in the semi-Dirac is linear in one direction, the quadratic behavior in the other direction seemingly dominates over the linear term which results in a similar characteristic of conductance quantization of a 2DEG.\par

\begin{figure}[]
\begin{center}
\subfloat[]{\includegraphics[width=0.25\textwidth]{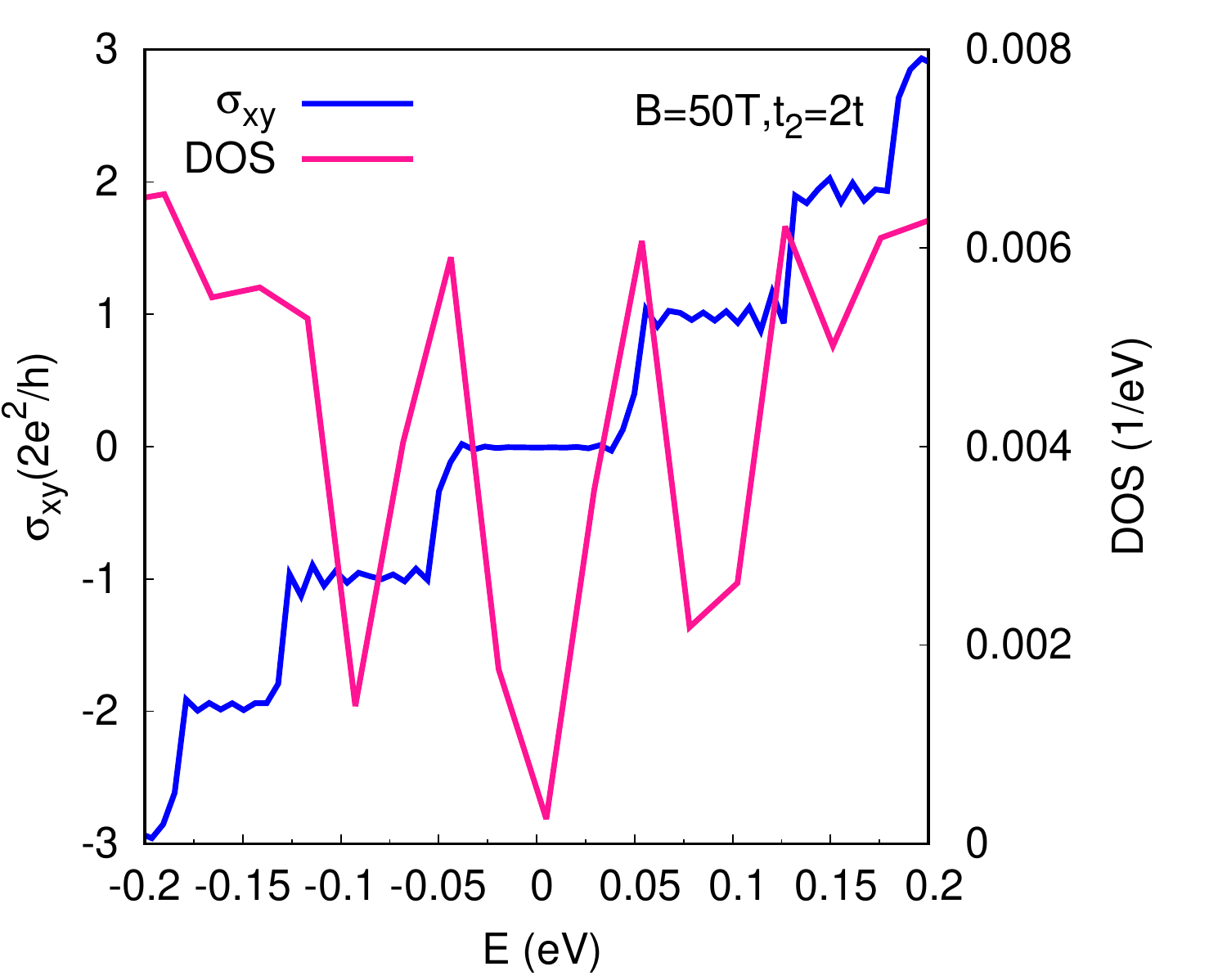}\label{fig:8a}} \hspace*{- 0.2 cm}
\subfloat[]{\includegraphics[width=0.25\textwidth]{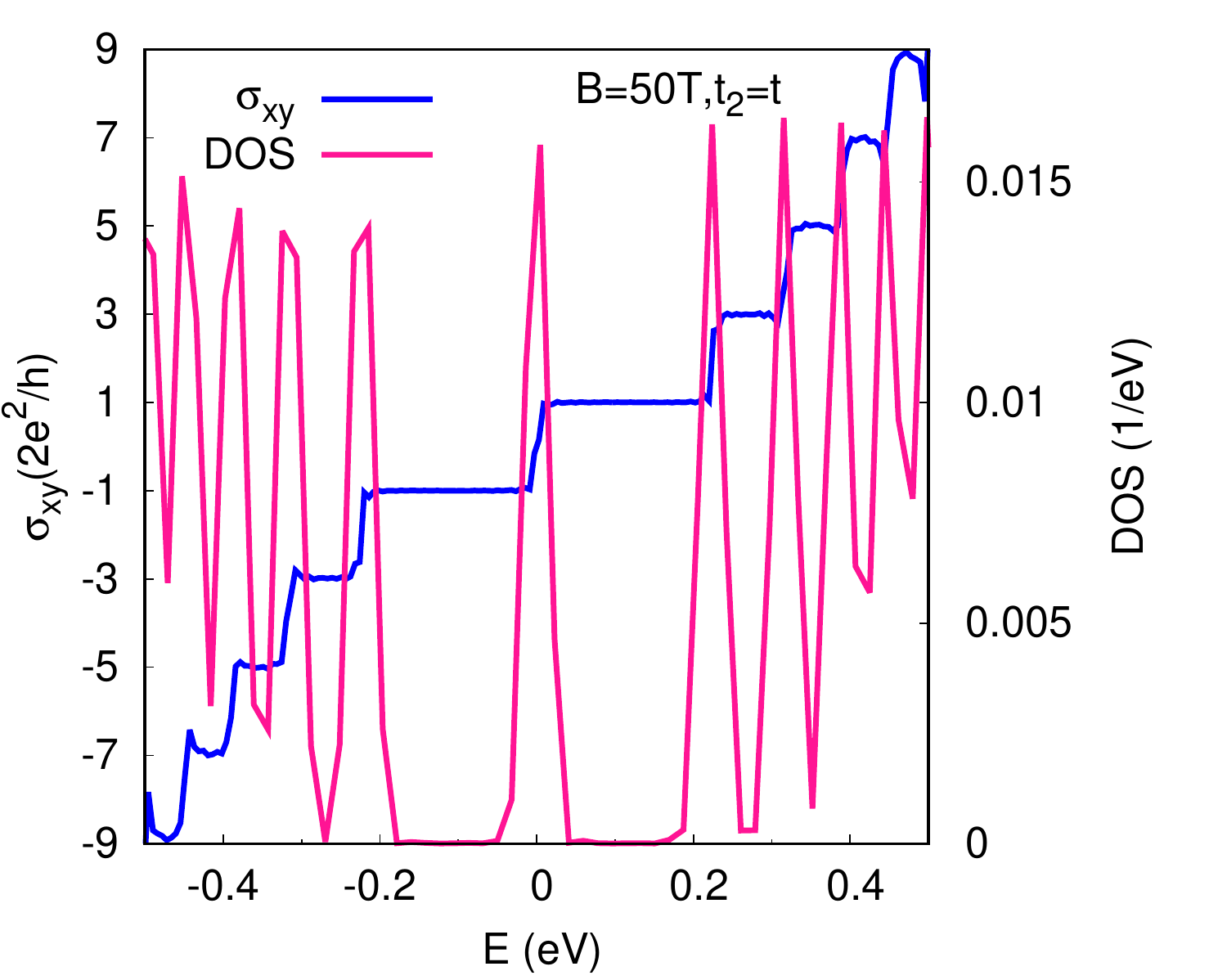}\label{fig:8b}}\\
\subfloat[]{\includegraphics[width=0.25\textwidth]{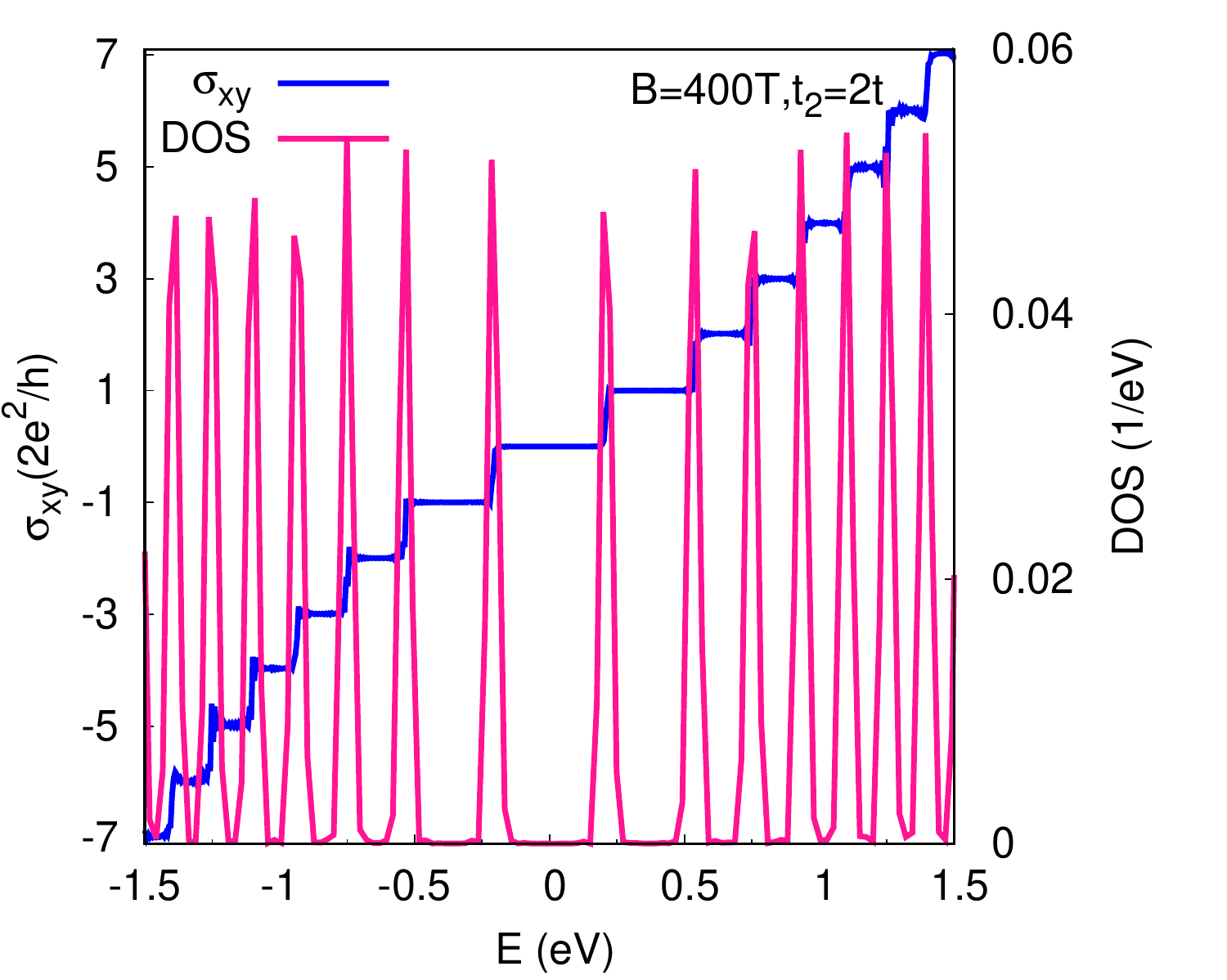}\label{fig:8c}} \hspace*{- 0.2 cm}
\subfloat[]{\includegraphics[width=0.25\textwidth]{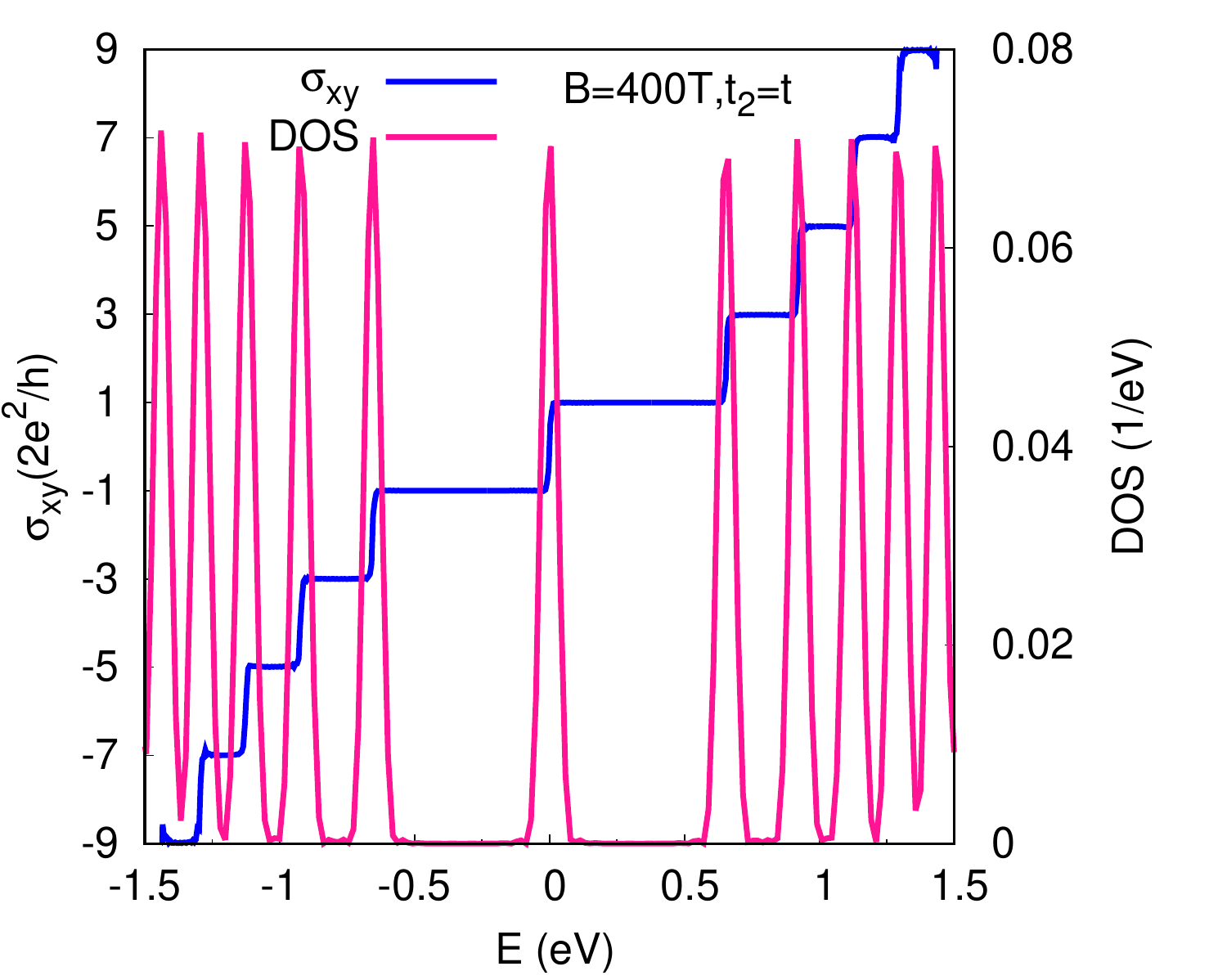}\label{fig:8d}}
\caption{(Color online) Hall conductivity, $\sigma_{xy}$ (in units of $2e^2/h$) and the DOS (in units of 1/eV) is plotted as a function of Fermi energy, $E$ (in units of eV) for different cases (a) $t_2$ = $2t$, $B=50$T (b) $t_2$ = $t$, $B=50$T (c) $t_2$ = $2t$, $B=400$T and (d) $t_2$ = $t$, $B=400$T.}
\label{fig:8}
\end{center}
\end{figure}

\begin{figure}[h]
\begin{center}
\subfloat[]{\includegraphics[trim=0 0 0 0,clip,width=0.28\textwidth,height=0.20\textwidth]{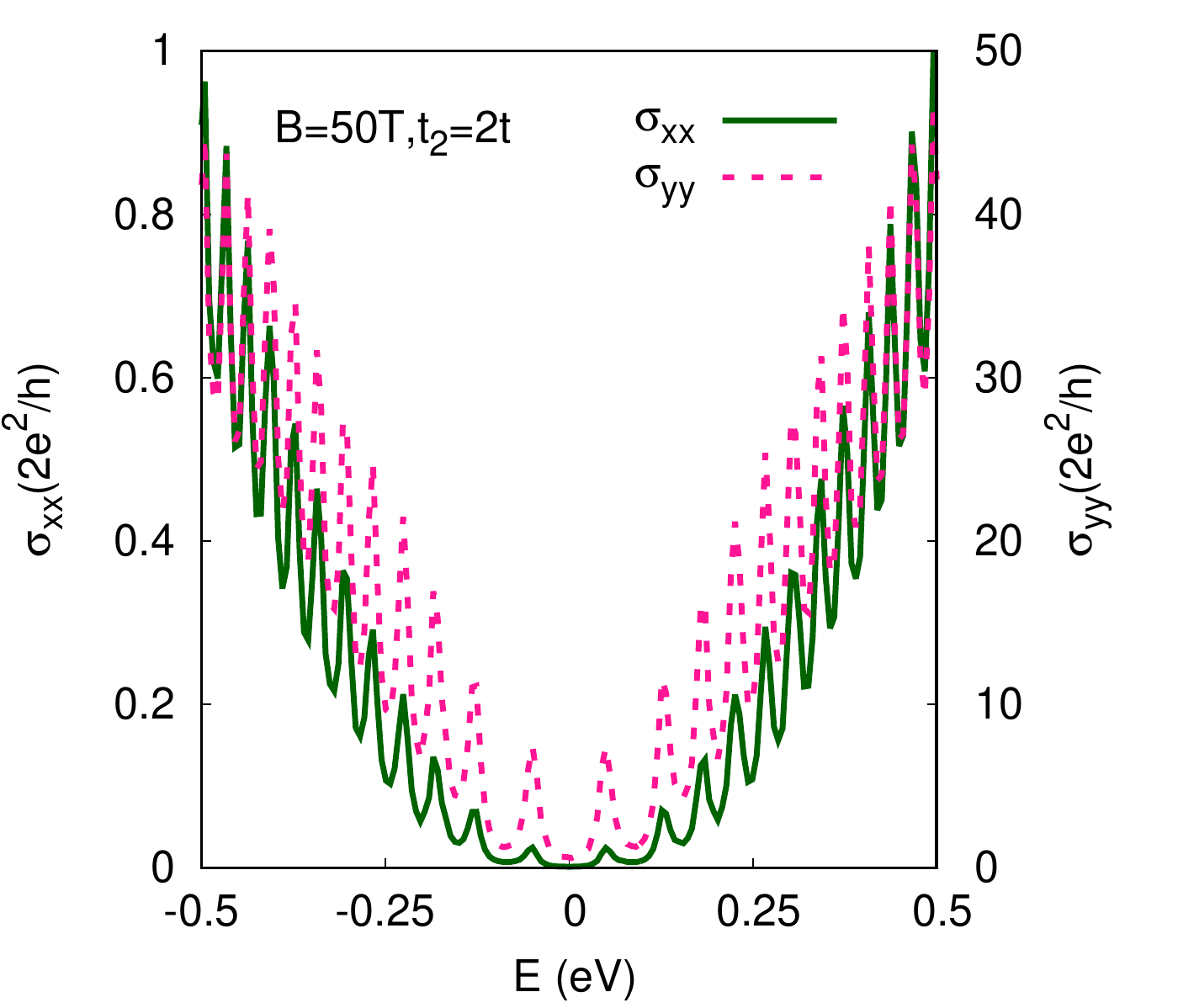}\label{fig:9a}} \hspace*{-0.3 cm}
\subfloat[]{\includegraphics[width=0.23\textwidth]{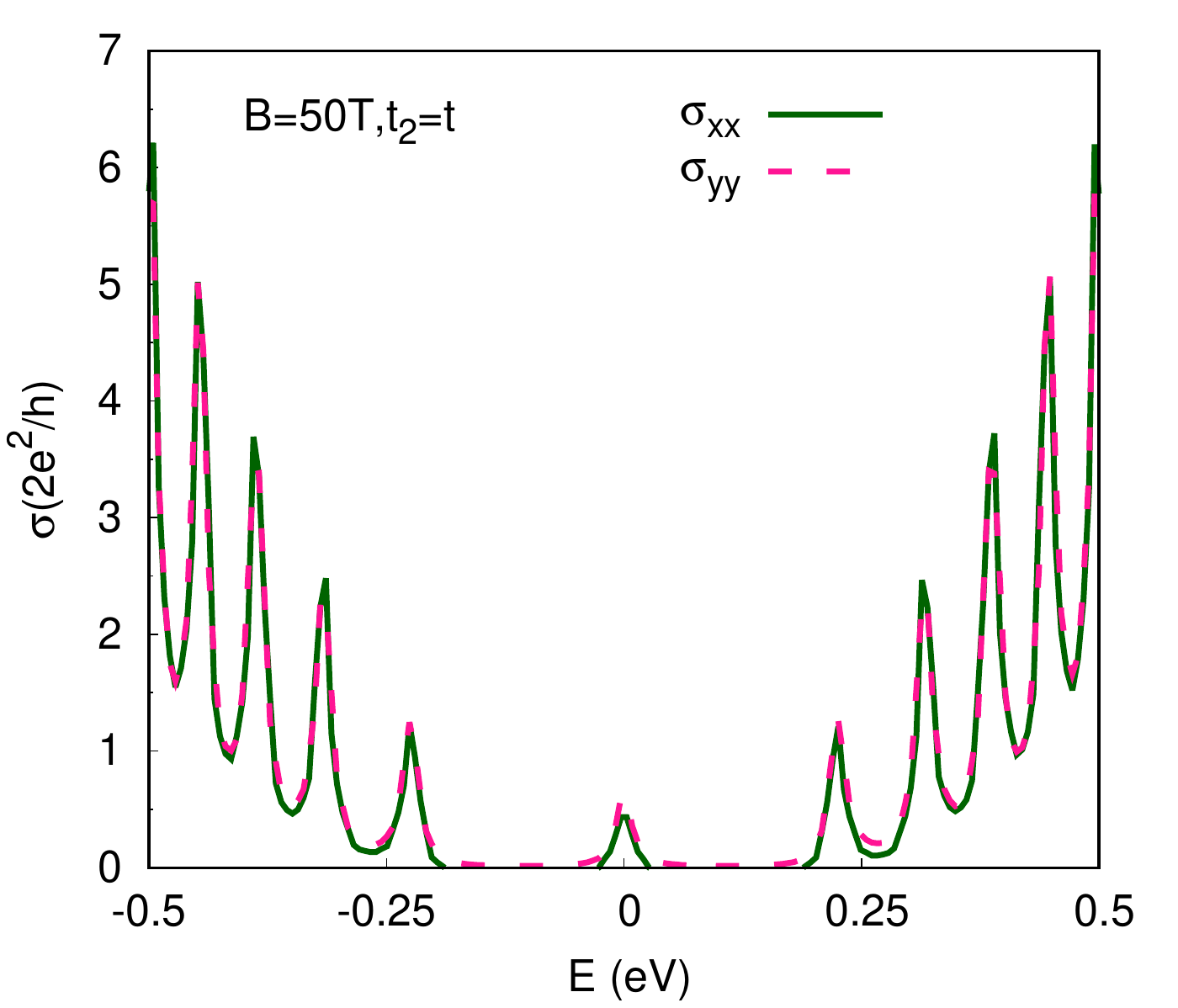}\label{fig:9b}}\\
\subfloat[]{\includegraphics[trim=0 0 0 0,clip,width=0.28\textwidth,height=0.20\textwidth]{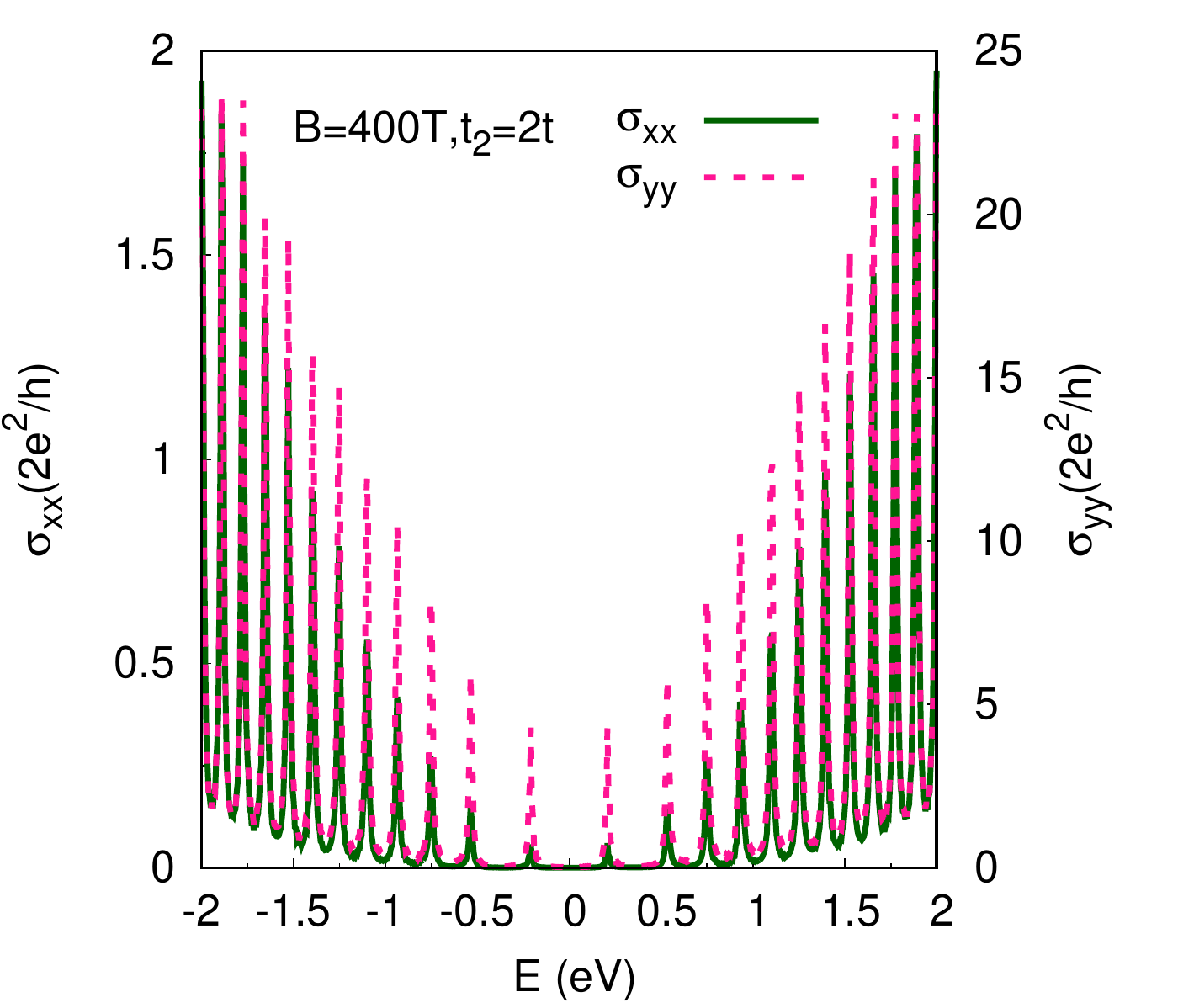}\label{fig:9c}} \hspace*{- 0.3 cm}
\subfloat[]{\includegraphics[width=0.23\textwidth]{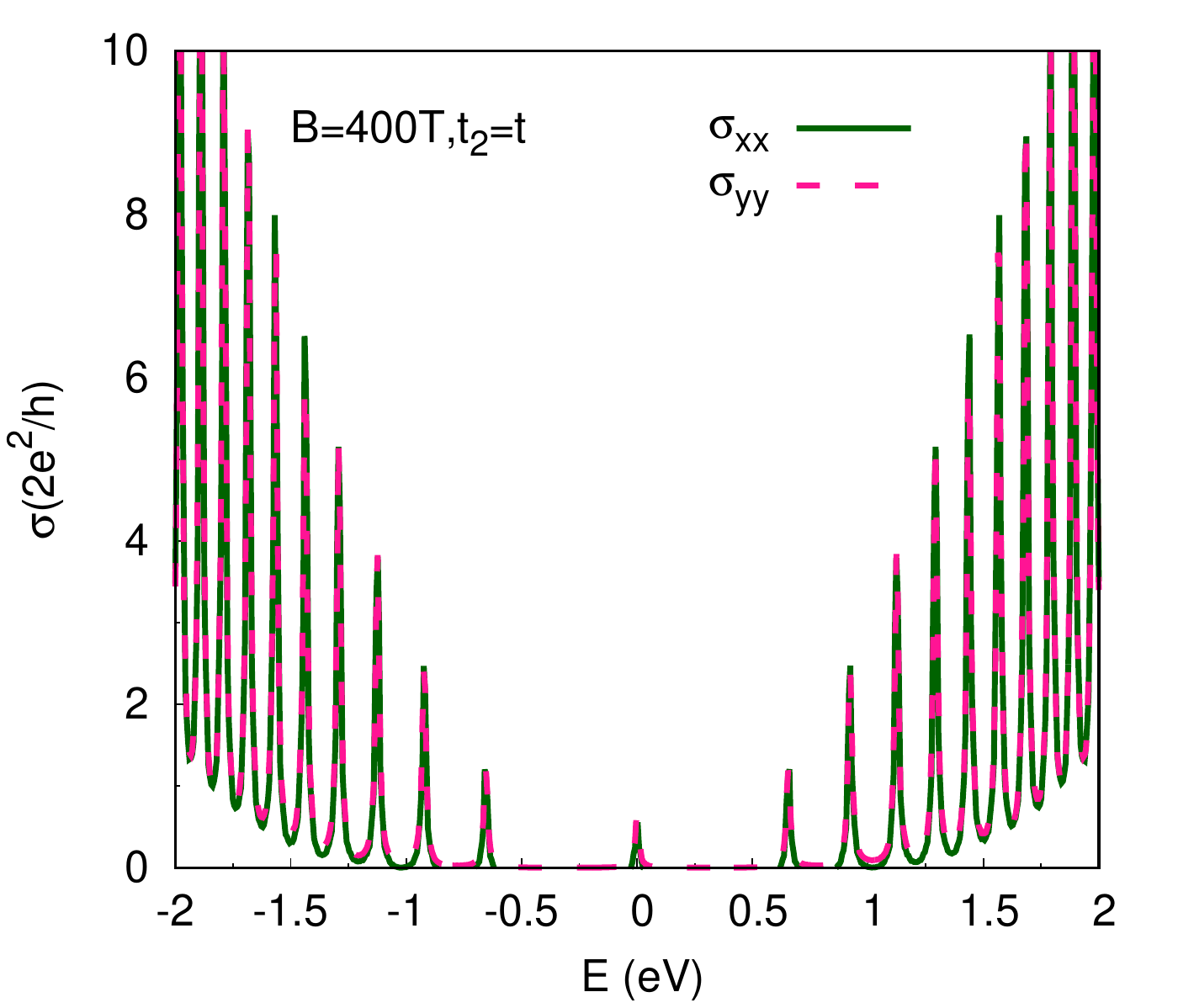}\label{fig:9d}}\\
\subfloat[]{\includegraphics[trim=0 0 0 0,clip,width=0.28\textwidth,height=0.20\textwidth]{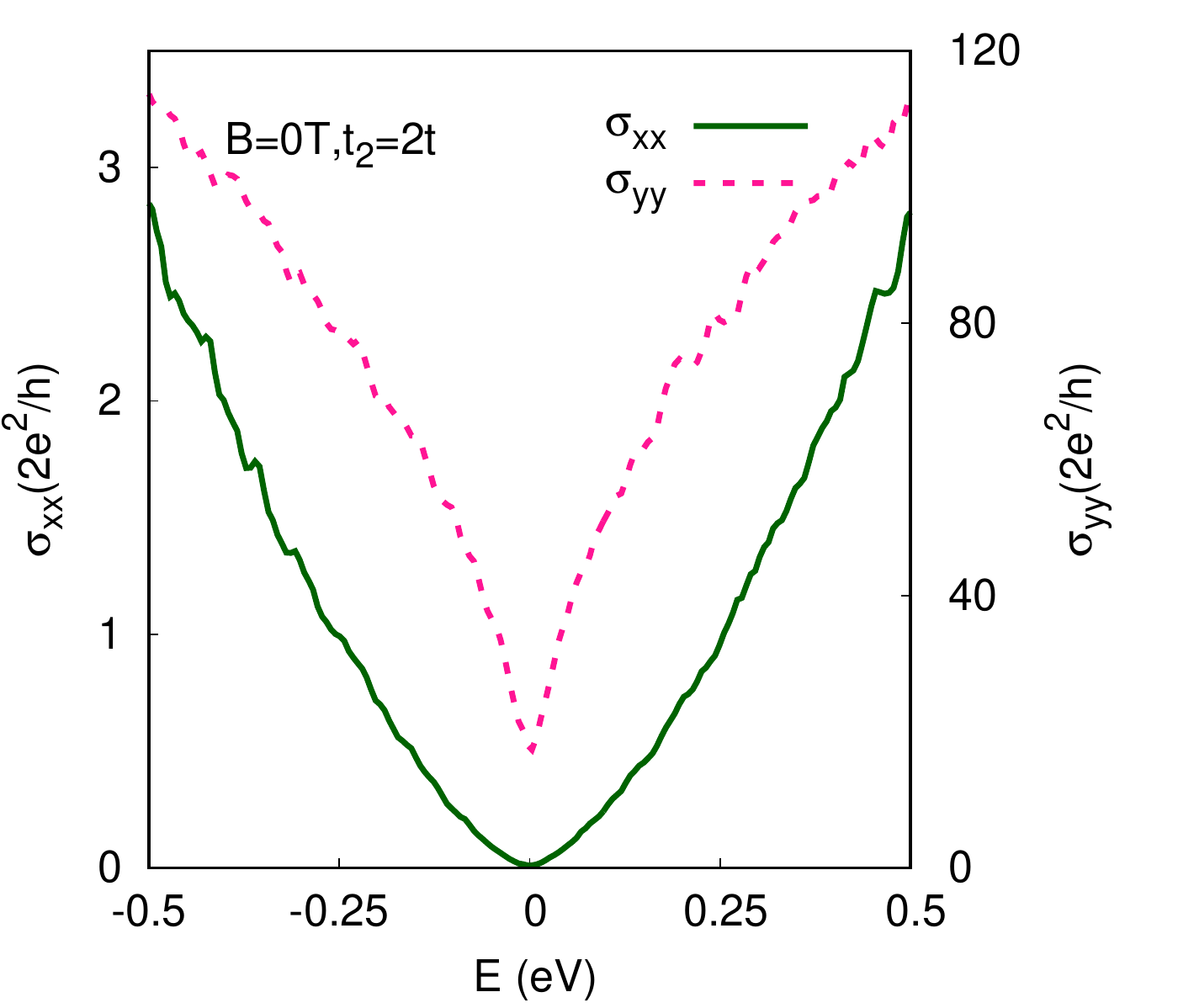}\label{fig:9e}} \hspace*{-0.3 cm}
\subfloat[]{\includegraphics[width=0.23\textwidth]{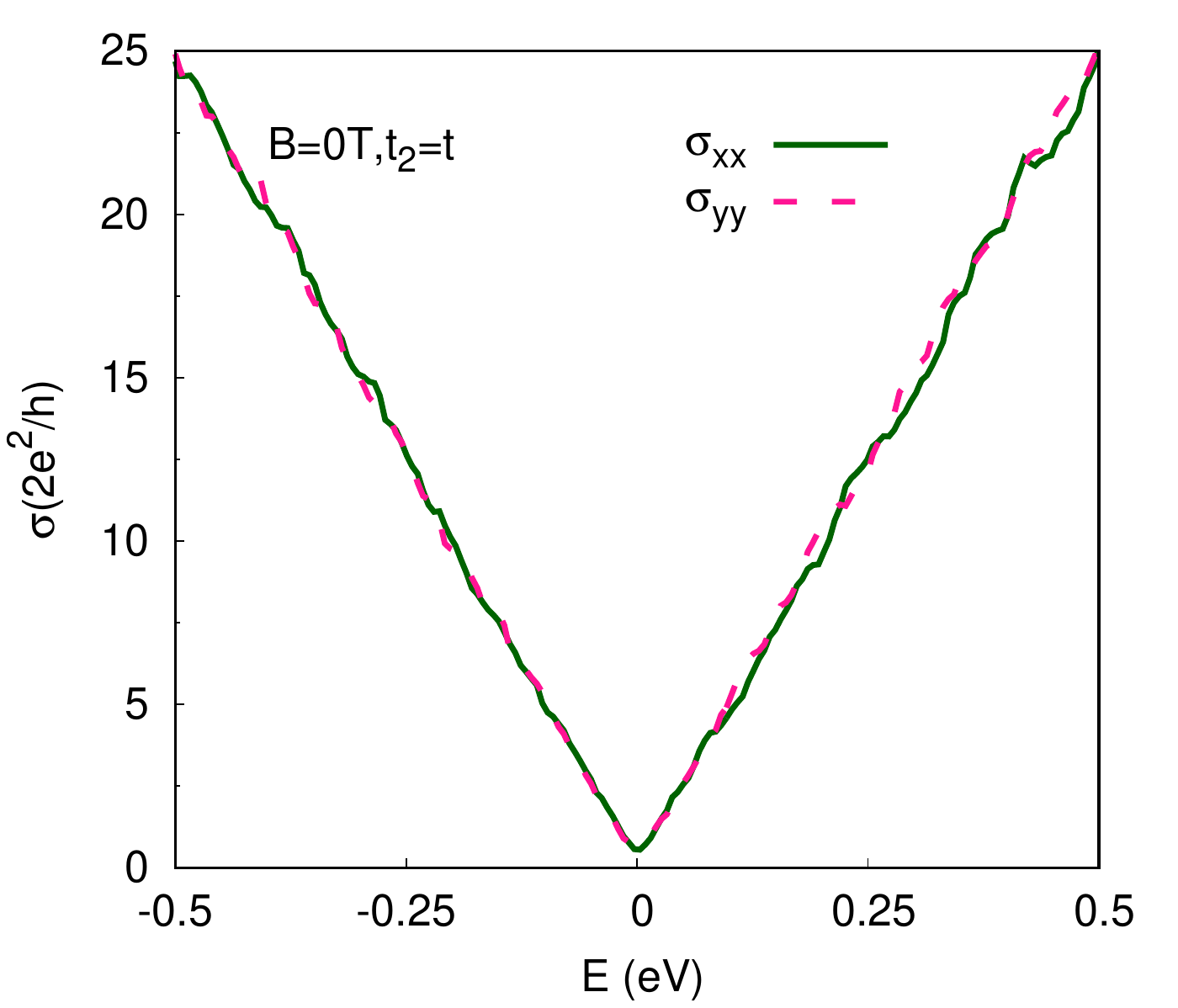}\label{fig:9f}}
  \caption{(Color online) Longitudinal conductivity, $\sigma_{xx}$ and $\sigma_{yy}$ (in units of $2e^2/h$) is plotted as a function of Fermi energy, $E$ (in units of eV) for different cases (a) $t_2$ = $2t$, $B=50$T (b) $t_2$ = $t$, $B=50$T (c) $t_2$ = $2t$, $B=400$T (d) $t_2$ = $t$, $B=400$T (e) $t_2$ = $2t$, $B=0$T and (f) $t_2$ = $t$, $B=0$T.}
\label{fig:9}
\end{center}
\end{figure}
In Fig.~\ref{fig:8}, we have shown the Hall conductivity, $\sigma_{xy}$ and the DOS for $B=50$T and $400$T in the same frame. In the presence of the magnetic field, the DOS consists of peaks of discrete energy levels (Landau levels) as shown in Fig.~\ref{fig:8}. The DOS vanishes in the plateau region and shows a sharp peak corresponding to a Landau level when the Hall conductivity goes through a transition from one plateau to another. However we get broad DOS peaks at lower values of the magnetic field ($B = 50$T) which is particularly visible for the semi-Dirac case owing to the small energy separation between the Landau levels (less than 3 meV). Sharper peaks will require computation of very large number of Chebyshev moments. Figure~\ref{fig:8a} shows that there is no Landau level peak at zero energy for $t_2=2t$ which is also characteristic of a conventional 2DEG in contrast to $t_2=t$ case, where the zero-energy peak is well-observed (see Fig.~\ref{fig:8b}). The presence of a zero-energy peak for the Dirac case is related to the chiral anomaly present there. Figures~\ref{fig:8c} and \ref{fig:8d} show that several Landau levels can be observed with the same qualitative behavior for very high magnetic field $B=400$T for both the semi-Dirac and the Dirac systems. The Hall conductance is quantized due to the quantized Landau level. It is interesting to note that although the energy does not depend linearly on the Landau level index, $n$ and magnetic field, $B$ in the case of a semi-Dirac system, as said earlier, the quantized value of $\sigma_{xy}$ of a semi-Dirac material is analogous to that of a 2DEG. It is once again pertinent to mention that the Landau level spectra in phosphorene in a perpendicular magnetic field depending on the index, $n$ (an additional factor of `2' will appear for spin degeneracy) has connections with the semi-Dirac physics \cite{chang,roldan}.\par

To investigate the magneto-transport, we further calculate the longitudinal conductivity along the $x$ ($\sigma_{xx}$) and $y$ ($\sigma_{yy}$) directions. Figures~\ref{fig:9a} and \ref{fig:9b} show the longitudinal conductivity, $\sigma_{xx}$ (green curve) and $\sigma_{yy}$ (magenta curve) as a function of the Fermi energy, $E$ for moderate values of the $B$ field $B$=50T for a semi-Dirac and a Dirac system respectively. The longitudinal conductivity reveals largely anisotropic behavior owing to the presence of anisotropy in the dispersion for $t_2=2t$. The component of $\sigma$ in the $x$ ($\sigma_{xx}$) and $y$ ($\sigma_{yy}$) directions are quantitatively different in nature. The magnitude of $\sigma_{yy}$ is larger than that of $\sigma_{xx}$. This is definitely a contrasting feature compared to the case of a Dirac system where the magnitudes of $\sigma_{xx}$ and $\sigma_{yy}$ are same as shown in Fig.~\ref{fig:9b}. Moreover, the absence of zero-energy peak also supports our discussion on the Landau level results (see Fig.~\ref{fig:8c}) for the semi-Dirac material. Figures~\ref{fig:9c} and \ref{fig:9d} show similar results for very high values of the magnetic field, namely, $B=400$T with the well-observed sharp peaks at larger values of energy. The amplitudes increase at large values of the Fermi energy owing to the increase in the scattering rate of the Landau levels as one goes to higher values of $n$ for both the semi-Dirac and the Dirac systems. Since the Landau levels are not equidistant for both of the cases, the interval between the peaks are not spaced equally. It can be seen that the longitudinal conductivity is non-vanishing only when the Fermi energy is within a Landau band where the backscattering processes are present. \par
To compare and contrast further between the two cases, we plot the longitudinal conductivities ($\sigma_{xx}$ and $\sigma_{yy}$) in the absence of any magnetic field ($B = 0$) in Figs.~\ref{fig:9e} and \ref{fig:9f} for the semi-Dirac and the Dirac systems respectively. Apart from the suppression of the conductivities by one order of magnitude by the magnetic field, one can take a note of the linear dependence of the conductivity on the Fermi energy, $E$ for the Dirac case\cite{novak}, while they appear with different exponents for the semi-Dirac one. The feature is qualitatively same as that observed for $B \neq 0$. However, the peaks in the conductance spectra vanish at $B = 0$ owing to the absence of Landau levels.  \par  
\section{Summary}\label{V}
In this work, we have studied the influence of a perpendicular magnetic field with a semi-Dirac dispersion within the framework of a tight-binding model of a honeycomb lattice. In order to compare and contrast with a prototype Dirac material, such as, graphene, we have presented our results for both cases. We consider a semi-Dirac nanoribbon with a finite width and study the Hofstadter butterfly and properties of the Landau level spectra. We have observed two identical gapped spectra symmetrically placed above and a below $E=0$ and along with a zero-energy mode in the Hofstadter butterfly spectrum in contrast to what is observed for a Dirac system. The Landau level becomes fully dispersive in the bulk for moderate values of the magnetic flux which is not true for the Dirac case. Furthermore, we explore the magneto-transport properties using the Kubo formula. We observed that the Hall conductivity shows standard quantization similar to that of a conventional semiconductor 2DEG with a parabolic band, which is highly contrasted with respect to a Dirac system. The zero Landau level peak is absent in the case of a semi-Dirac system. The longitudinal conductivities, $\sigma_{xx}$ and $\sigma_{yy}$ show anisotropic behaviour due to the distinct dispersion in two longitudinal directions.    

\section{ACKNOWLEDGMENTS} 
PS acknowledges Sim$\tilde{a}$o M. Jo$\tilde{a}$o, SK Firoz Islam, Abbout Adel for useful discussions and Rafiqul Rahaman for helping in computation. SB thanks SERB, India for financial support under Grant No. EMR/2015/001039. SM is supported by JSPS KAKENHI under Grant No. JP18H03678.

\end{document}